\documentclass[aps]{revtex4}
\usepackage{amsmath,amssymb,epsfig}
\usepackage{graphicx}

\begin{document}
\title{Brane-cosmology dynamics with induced gravity}
\author{Burin Gumjudpai}

\address{ Institute of Cosmology \& Gravitation, University of Portsmouth,
Portsmouth, PO1 2EG, UK}

\begin{abstract}
The Friedmann equations for a brane with induced gravity are
analyzed and compared with the standard general relativity and
Randall-Sundrum cases. Randall-Sundrum gravity modifies the early
universe dynamics, whereas induced gravity changes the late
universe evolution. The early and late time limits are
investigated. Induced gravity effects can contribute to
late-universe acceleration. The conditions for this are found.
Qualitative analysis is given for a range of scalar field
potentials.
\end{abstract}

\maketitle

\section{Introduction}

Recently, the idea of the universe as 3+1-dimensional braneworld
embedded in extra dimensions (called the bulk) has been proposed
in order to solve the hierarchy problem. Matter particles live on
this brane but only gravitons can travel out to the bulk. Due to
the leaking of gravity into the extra dimensions, one observes
higher dimensional gravity on small scales. The scale could be as
large as sub-millimeter and the fundamental scale of gravity could
be as low as 1 TeV \cite{ADD}. In the Randall-Sundrum (RS) II
scenario \cite{rspaper2}, our universe is a positive tension
brane. There is one non-compact extra dimension warped by a
negative bulk cosmological constant, $\Lambda_5$. Hence the bulk
metric is $AdS_5$. From the effect of warped geometry, the
fundamental gravity scale, $M_5$ is below the four-dimensional
Planck scale, $M_4$. A modified braneworld scenario was proposed
by Dvali, Gabadadze and Porrati (DGP) \cite{Dvali:2000hr} (see
also \cite{Collins:2000yb}). The key idea of the DGP model is the
inclusion of a four-dimensional Ricci scalar into the action. The
conventional four-dimensional gravity is recovered on scales
smaller than a crossover scale, $r_c$. Beyond this scale gravity
becomes five dimensional and weaker due to the suppression of this
term. The scale $r_c$ could be as large as the Hubble radius. The
merit of the DGP model is to explain late-time acceleration
\cite{Defy01}.

 We investigate the modification of solutions to
 the Friedmann equations in braneworld scenarios in comparison to the
standard cosmology, especially in the DGP case, and including the
case where the universe is dominated by a scalar field.  The
scalar field phase space with potentials that are of interest in
quintessential models, inflation and cold dark matter will be
illustrated in the contexts of standard cosmology, RSII and DGP
braneworlds.

\section{Basic Equations}
In braneworld scenarios, the gravity model is modified. However,
on our brane universe, cosmology should be consistent with the
standard FRW model. The standard Friedmann equation is
\begin{equation}
H^2 + \frac{k}{a^2} \,=\, \kappa^2 \frac{\rho}{3} +
\frac{\Lambda}{3}
\end{equation}
where $\kappa^2 = M_{4}^{-2} = 8\pi G$. Nucleosynthesis already
imposes the strong constraint that the standard Friedmann equation
should govern the expansion after $\rho \sim $ 1 MeV. Therefore at
this energy scale, the brane-modified Friedmann equations should
effectively approach the standard Friedmann equation. Table-top
gravity experiments tell us that the 4D Newton law is still valid
at least down to $0.1$ mm, hence 4D gravity can only be modified
for distances $l < 0.1$ mm.

On the brane in both RSII and DGP models the conservation of
energy and momentum holds:
\begin{equation}
\dot{\rho} = -3H(\rho+p) \label{flu}
\end{equation}

\subsection{Randall-Sundrum II braneworld}
In the RSII braneworld, the 5D Einstein equation in the bulk is
$^{(5)}G_{AB} = - \Lambda_{5}\, ^{(5)}g_{AB}$. Projecting the 5D
curvature and imposing Darmois-Israel junction conditions and
$Z_2$ symmetry, Shiromizu et al. \cite{shiromizu} have found the
effective Einstein equation on the brane
\begin{equation}
G_{ab} = - \Lambda g_{ab} + \kappa^2 T_{ab} + 6
\frac{\kappa^2}{\lambda} {\mathcal{S}}_{ab} - {\mathcal{E}}_{ab}
\end{equation}
where ${\mathcal{S}}_{ab}$ is the high-energy correction term
which is quadratic in the brane energy momentum tensor,
${\mathcal{E}}_{ab}$ is the projected bulk Weyl tensor on the
brane and $\lambda$ is the brane tension. Assuming that the brane
is FRW, the RSII Friedmann equation on the brane is \cite{bit}
\begin{equation}
H^2 + \frac{k}{a^2} \,=\, \kappa^2 \frac{\rho}{3}\big( 1 + \frac
{\rho}{2\lambda}\big) + \frac{m}{a^4} + \frac{\Lambda}{3}
\label{frrs}
\end{equation}
where $m$ is a constant of integration obtained from the bulk Weyl
tensor. The $m/a^4$ term is called dark radiation. We can
fine-tune the cosmological constant on the brane by the relation
\begin{equation}
\Lambda = \frac{1}{2}(\Lambda_{5} + \kappa^2 \lambda)
\end{equation}
The negative bulk cosmological constant defines the $AdS_{5}$
curvature scale $l$ via $ \Lambda_{5} = - 6/l^2 $. The curvature
scale and $\kappa_5$ relate to each other via $ 1/l = \kappa_{5}^2
\lambda/6 $.  The relation between the 5D Planck mass, brane
tension and 4D Planck mass is $M_{5}^3 = M_{4} \sqrt{\lambda/6} $.
In this paper, we consider only scalar fields that live on the
brane; the RS braneworld with bulk scalar field has been discussed
\cite{davidl}.

\subsection{Dvali-Gabadadze-Porrati braneworld}
In the DGP braneworld, unlike the RSII model, the infinite extra
dimension is flat ($\Lambda_{5} = 0$). Brane tension is assumed to
be zero or cancelled out with a brane cosmological constant. (The
model can be generalized to include nonzero $\Lambda_5$ and
$\lambda$ (see
Ref.\cite{Collins:2000yb,Kiritsis:2002ca,Sahni:2002vs,Maeda:2003ar})
but we do not consider this case here.)  The model considers
effects of induced gravity on the brane through the one-loop
process of graviton and brane matter interaction. Induced gravity
yields a Ricci scalar term in the Einstein action
\cite{Dvali:2000hr}. The 5D Einstein equation has the form
\begin{equation}
^{(5)}G_{AB} \equiv  \, ^{(5)}R_{AB} - \frac{1}{2}\, ^{(5)}R \:
^{(5)}g_{AB} = \kappa_{5}^2 \Big[ \,^{(5)}T_{AB} + \,^{(5)}U_{AB}
\Big]
\end{equation}
where $\,^{(5)}U_{AB}$ is a contribution of the induced brane
scalar curvature. In the DGP model we have the usual potential
$V(r)\sim 1/r $ at small scales while gravity becomes 5D at scales
larger than a crossover scale
\begin{equation}
r_c \equiv \frac{M_{4}^2} {2M_{5}^3}
\end{equation}
where the potential transforms to $V(r)\sim 1/r^{2}$. The DGP
Friedmann equation is
\begin{equation}
H^2 + \frac{k}{a^2}\,=\,
 \Bigg(\sqrt{ \frac{\kappa^2 \rho}{3} + \frac{1}{4r_{c}^{2}} }  + \frac{\epsilon}{2r_{c}}
 \Bigg)^2  \label{DGPfr}
\end{equation}
where $\epsilon = \pm 1$ gives two branches of solutions
\cite{Defy01}. The signs correspond to how the brane is embedded
into the bulk \cite{ruth}. Cosmological implications of the DGP
braneworld have been analyzed \cite{DDG,bbscan, Dick:2001np}.

\section{Solution of Friedmann Equation}
From now on we assume that the universe is flat with negligible
cosmological constant. For the RSII Friedmann equation with $k =
\Lambda = m = 0$ and constant $w$, the exact solution is
\cite{bit}

\begin{equation}
a \;=\; {\mbox{const}} \left[ t(t + t_{\lambda})
\right]^{1/3(1+w)},\;\:\;\;  t_{\lambda} = \frac{\sqrt{6}}{2
\sqrt{\lambda}} M_4 < 10^{-9} {\mbox{sec}} \label{rsscalef}
\end{equation}
where $w \neq -1$. If we include the dark radiation, the solution
for the radiation era is \cite{barrowmaartens}
\begin{equation}
a \;=\; {\mbox{const}} \left[ t(t + t_{\lambda})
\right]^{1/4},\;\:\;\; t_{\lambda} = \frac{\sqrt{6} M_4}{2 \sqrt{
\lambda}(1+\rho^{*}/\rho)}
\end{equation}
These solutions recover the standard solution in the radiation era
when $t \gg t_{\lambda}$, as $a \propto t^{1/2}$. The solution is
different from standard cosmology when $t \ll t_{\lambda}$ and it
is $a \propto t^{1/4}$. Here we will solve the DGP Friedmann
equation in the high-energy and low-energy regimes of the
universe.

\subsection{Solution in the late universe}
The DGP Friedmann equation (\ref{DGPfr}) with $k=0$ can be
rewritten as

\begin{eqnarray}
H^2\: &=& \:\frac{1}{4r_{c}^2} \Bigg[ \sqrt{1 + \frac{4\rho
r_{c}^2}{3M_{4}^2}} + \epsilon  \Bigg]^2 \nonumber\\
   \: &=&\: \frac{1}{4r_{c}^2} \Bigg[ \bigg( 1+ \frac{4\rho r_{c}^2}{3M_{4}^2} \bigg)
    + 2 \epsilon \sqrt{1+ \frac{4\rho r_{c}^2}{3M_{4}^2}} +
   1
   \Bigg]\label{dgpbi}
\end{eqnarray}
We solve the above equation by expanding it in terms of $\rho
r_c^2/M_4^2 \ll 1$. At zeroth order, the equation becomes
\begin{equation}
H^2 \rightarrow  \frac{1}{2r_{c}^2} \left( 1 + \epsilon \right)
\end{equation}
If $\epsilon = +1$, $ H^2 \rightarrow 1/r_{c}^2$ in agreement with
\cite{Defy01}. This yields $a \sim \exp(r_{c}^{-1} t)$, i.e.
accelerating late-universe expansion. In the case $\epsilon = -1
$, $ H^2 \rightarrow 0$, implying that $ a \sim$ constant, i.e.
the universe is asymptotically static. At the next order, we
consider the two branches of solution separately.

\begin{itemize}
\item
In the case of $ \epsilon = +1$ we obtain

\begin{equation}
H^2 \; \simeq \;  \frac{1}{r_{c}^2} + \frac{2\rho}{3M_{4}^2}
\label{eppos1}
\end{equation}

Assuming matter domination $\rho = \rho_{0}(a_{0}/a)^3$, the
solution is
\begin{equation}
a \simeq  a_{0} \Big(\frac{2 \rho_{0} r_{c}^2}{3 M_{4}^2}
\Big)^{1/3} \sinh^{2/3}\left[\frac{3}{2 r_{c}}(t-\tau)\right]
\end{equation}
where $\tau$ is an arbitrary constant of integration. At very late
times, extra-dimension effects dominate the expansion completely,
and the expansion accelerates as $ a \simeq a_{0} \left[  \rho_{0}
r_{c}^2 /(6 M_{4}^2) \right]^{1/3} \exp{[(t-\tau)/r_c]}$.

\item When $ \epsilon = -1$, we obtain at lowest order
\begin{equation}
 H \;  \simeq  \; \frac{\rho r_{c}}{3 M_{4}^2} \label{epneg1}
\end{equation}
with solution
\begin{equation}
a \simeq  a_{0} \left[3(1+w) \frac{\rho_0 r_c}{3 M_4^2}
\right]^{1/3(1+w)}  \: (t-\tau)^{1/3(1+w)}   \label{scaleepneg}
\end{equation}
For matter domination at late times, we obtain $a \propto
t^{1/3}$. This solution implies slower expansion than the standard
cosmology for which $a \propto t^{2/3}$. In this case, the late
time DGP brane universe does not give acceleration and one needs
dark energy to dominate in order to obtain acceleration at late
time. Note that equation (\ref{epneg1}) looks similar to the RSII
Friedmann equation when the quadratic term dominates at high
energy. Therefore it is not surprising that the expansion of this
case is similar to equation (\ref{rsscalef}) of the RSII model.
\end{itemize}

\subsection{Solution in the early universe}

We rearrange the Friedmann equation (\ref{DGPfr}) with $k=0$ as
\begin{equation}
H^2 \:=\:  \frac{\rho}{3M_4^2} \left[ \sqrt{1 + \frac{3M_4^2}{4
\rho r_c^2}} +  \frac{\epsilon}{2r_c}\sqrt{\frac{3
M_4^2}{\rho}}\right]^2
\end{equation}
At high energy $\rho r_c^2/M_4^2 \gg 1$, we can expand this as
\begin{equation}
H^2 \;=\; \frac{\rho}{3M_4^2} \left[\: 1 + \frac{1}{2}
\left(\frac{3 M_4^2 }{4 \rho r_{c}^2}\right) + \;\;\: \cdots
\;\;\: + \frac{\epsilon}{2r_c}\sqrt{\frac{3 M_4^2}{\rho}}
\:\right]^2
\end{equation}

To zeroth order, we obtain the standard cosmology Friedmann
equation $H^2 \simeq \rho/ 3M_{4}^2$. However if we include the
$\epsilon/r_c $ term,
\begin{equation}
H^2 \; \simeq \; \frac{\rho}{3 M_{4}^2} \left(\: 1 +
\frac{\epsilon}{ r_c} \sqrt {\frac{3M_{4}^2}{ \rho}}  \: \right)
\label{early}
\end{equation}
Assuming radiation domination, this can be solved to obtain
\begin{equation}
a \simeq a_{0} \left( \frac{4 \rho_{0}}{3 M_{4}^2 } \right)^{1/4}
(t-\tau)^{1/2} \left[ 1 + \frac{\epsilon}{4 r_c}(t-\tau) \right]
\end{equation}
where the second term in square brackets is the correction from
the extra dimension to the standard evolution.

\section{Condition for acceleration}

The condition for accelerating expansion is

\begin{eqnarray}
\dot{H} + H^2 \: = \: \frac{\ddot{a}}{a}\: > \:0
\end{eqnarray}

After differentiating $H^2$ with respect to time and using
equation (\ref{flu}), the acceleration condition for standard
cosmology takes the form:
\begin{equation}
\dot{H} + H^2 \:= \:-\frac{\kappa^2}{6}\big( \rho + 3p  \big)\:
> \:0
\end{equation}
and this implies $ p  < -\rho/3 $. In the braneworld cosmologies,
the conditions are modified.

\subsection{Condition for acceleration in RSII brane cosmology}
During an inflationary phase, the dark radiation term of the RSII
model is diluted away very quickly and can be considered
negligible. By the same procedure as in standard cosmology, one
can find the RSII acceleration condition as
\begin{equation}
\dot{H} + H^2 \:=\: -\frac{\kappa^2}{6} \left[ \rho\left(1+
\frac{2\rho}{\lambda}\right) + 3p \left(1+
\frac{\rho}{\lambda}\right) \right] \:
> \: 0
\end{equation}
which implies \cite{mwbh}
\begin{equation}
p \: <  \:-\frac{\rho}{3} \Bigg[ \frac{1+2\rho/\lambda}{1+
\rho/\lambda} \Bigg] \label{rsac}
\end{equation}
At low energies, $ \rho/\lambda \ll 1$, hence
\begin{equation}
p \:   \lesssim  \: -\frac{\rho}{3}\:
\Big(1+\frac{2\rho}{\lambda}\Big) \Big(1 -
\frac{\rho}{\lambda}\Big) \:  \simeq  \: -\frac{\rho}{3}\: \Big( 1
+ \frac{\rho}{\lambda}\Big)
\end{equation}
showing the correction to standard cosmology. On the other hand at
high energies, $\rho/\lambda \gg 1$, the condition becomes
\begin{equation}
p \: \lesssim  -\frac{\rho}{3}\: \Big(\frac{\lambda}{\rho}+2\Big)
\Big(1 - \frac{\lambda}{\rho}\Big)  \:  \simeq \:
-\frac{\rho}{3}\Big(2- \frac{\lambda}{\rho}\Big)
\end{equation}
or $ p \lesssim - 2\rho/3$. Hence it modifies the standard
cosmology condition significantly at high energies.

\subsection{Condition for acceleration in DGP brane cosmology}
By a similar method, the acceleration condition for DGP brane
cosmology is
\begin{equation}
\dot{H} + H^2 \:= \:-\frac{\kappa^2}{6}(\rho + p) \left[ 1+
\left(\kappa^2 \frac{\rho}{3}   + \frac{1}{4r_{c}^2}\right)^{-1/2}
\frac{\epsilon}{2r_{c}} \right] + \left[ \sqrt{ \kappa^2
\frac{\rho}{3} + \frac{1}{4r_{c}^2}  } + \frac{\epsilon}{2r_{c}}
\right]^2  \: > \: 0
\end{equation}
which implies
\begin{equation}
p \: < \:  - \rho \: +  \frac{2}{\kappa^2} \left[  \left(
\sqrt{\kappa^2 \frac{\rho}{3}+\frac{1}{4r_{c}^2}}
+\frac{\epsilon}{2r_{c}} \right)^2  \left(1+ \left[\kappa^2
\frac{\rho}{3}   + \frac{1}{4r_{c}^2}\right]^{-1/2}
\frac{\epsilon}{2r_{c}}\right)^{-1} \right]\:   \label{dgpac}
\end{equation}
At high energy, the $1/r_c$ term is small compared to the density
term and one can obtain the standard cosmology condition $p <
-\rho/3$. In the late universe, the extra-dimension effect can not
be neglected. We will use the approximation of the DGP Friedmann
equation previously derived instead of using the full form. For
the case $\epsilon = +1$, using the first-order late-time DGP
Friedmann equation (\ref{eppos1}), the acceleration condition
becomes
\begin{eqnarray}
\dot{H} + H^2 \:&  \simeq & \:  -\kappa^2(\rho + p ) \,+\,
\frac{1}{r_{c}^2}\, +\,
\frac{2\kappa^2 \rho}{3} \:>\: 0 \\
{\mbox{or}}\;\;\;\;\;\;\;\;\;\;\;\;\;\;\;\;\; p \;&<& \: -
\frac{\rho}{3} + \frac{1}{r_{c}^2 \kappa^2} \label{ppp1}
\end{eqnarray}
In the case $\epsilon = -1$, using equation (\ref{epneg1}), the
condition is simply
\begin{eqnarray}
 \dot{H} + H^2  \: & \simeq &\: -\frac{\kappa^4 r_{c}^2}{3}\,\rho(\rho+p) \, + \, \frac{\kappa^4\rho^2 r_{c}^2}{9}     \:> \:0 \\
{\mbox{or}}\;\;\;\;\;\;\;\;\;\;\;\;\;\;\;\;\; p \;&<& \: - \frac{2
\rho}{3} \label{ppp2}
\end{eqnarray}
These acceleration conditions will be used for analyzing the DGP
brane universe in next section when a scalar field is the dominant
component in the universe.

\begin{figure}[t]
\includegraphics[width=6cm,height=5cm,angle=0]{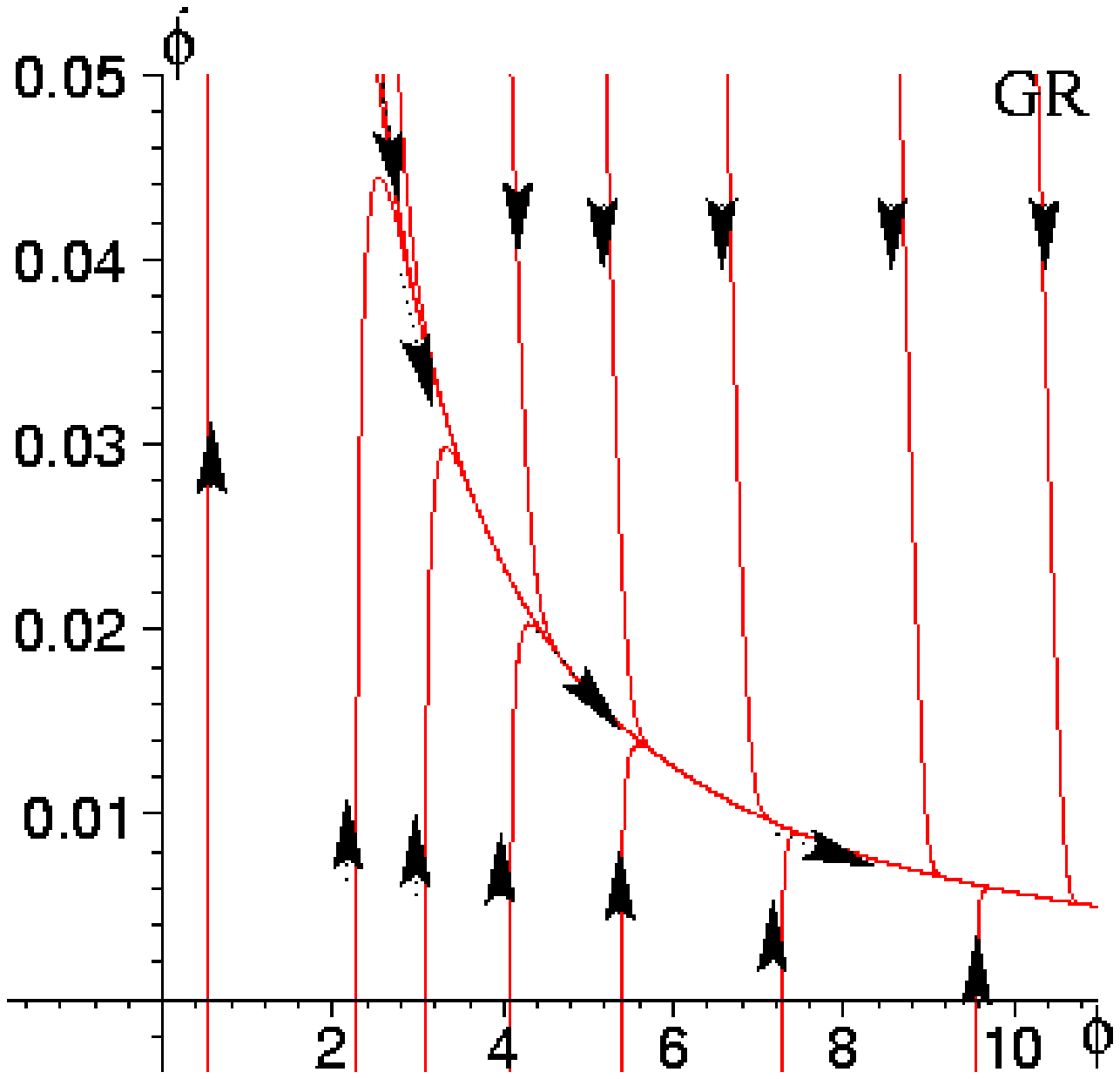}
\;\;\;\;\;\;\;\;\;\;
\includegraphics[width=6cm,height=5cm,angle=0]{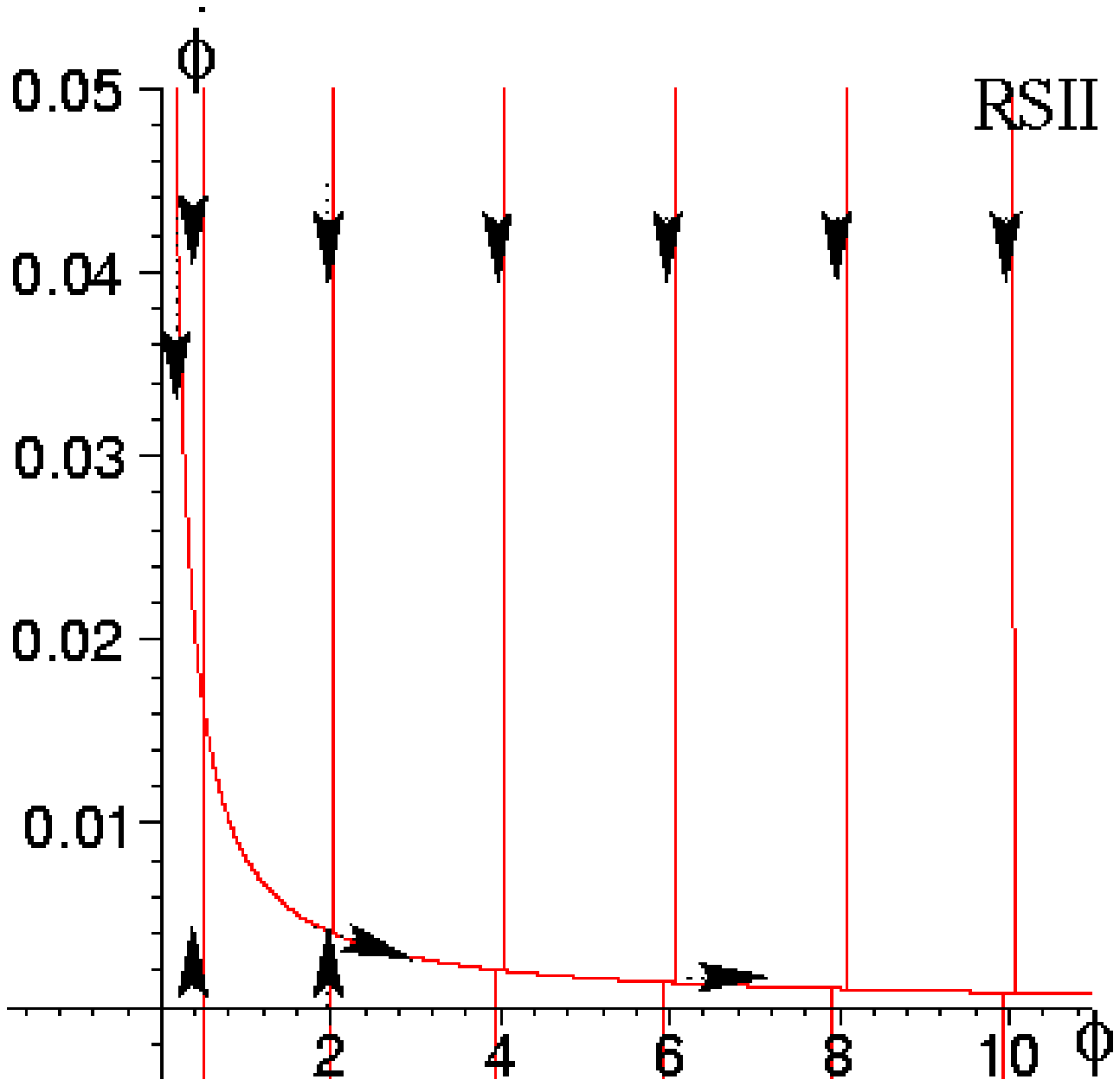}\\
\vspace{0.7cm}
\includegraphics[width=6cm,height=5cm,angle=0]{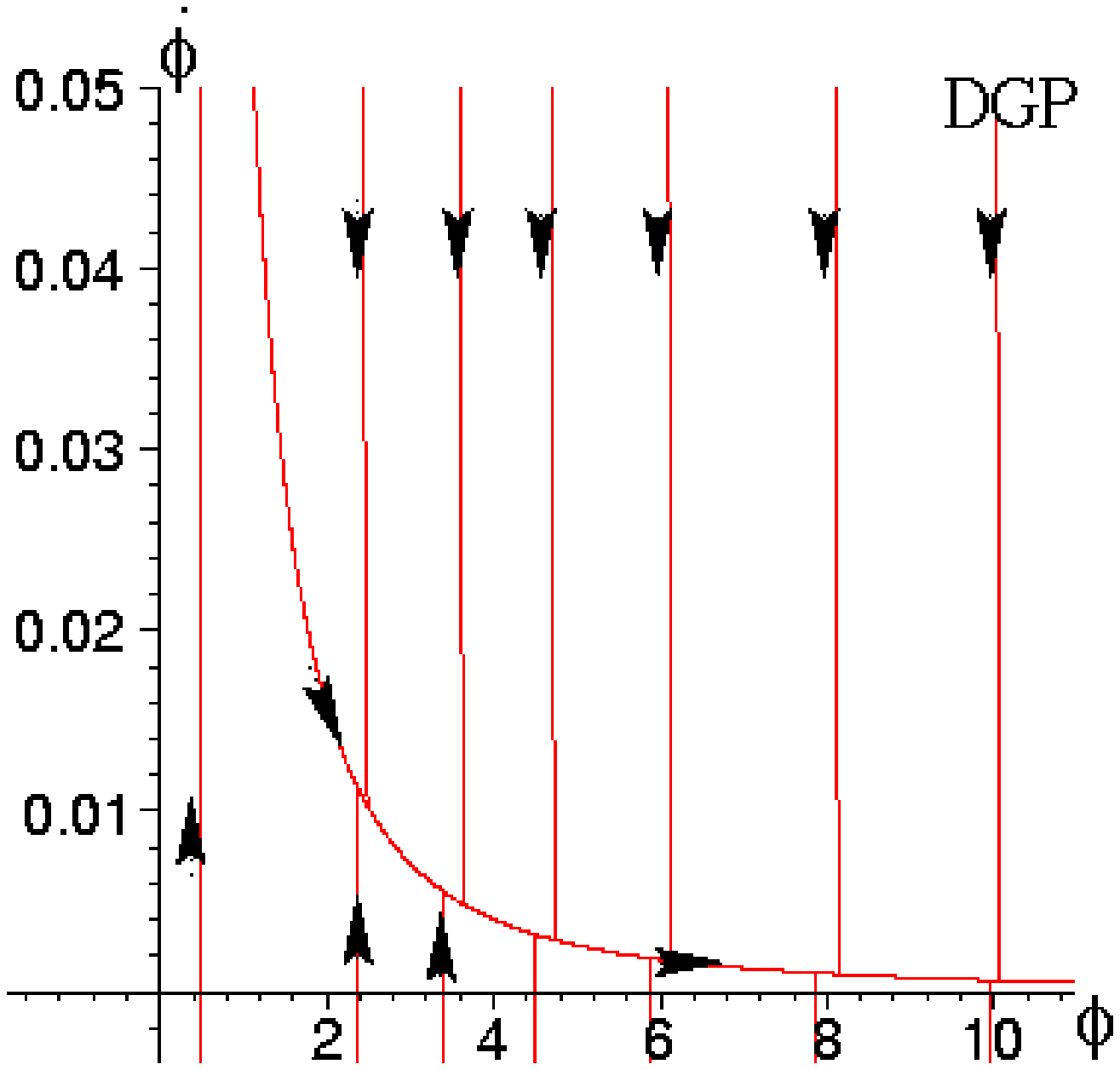}
 \caption{The phase space of a scalar field with $V(\phi)=\mu^{n+4} / {\phi^n}$, $n=1$, for the
standard cosmology, RSII and DGP brane cosmologies. We use $
\kappa=1 $, $\lambda = 10^{-4}$, $r_c = 2$, $ \mu^{n+4}=0.1 $ and
$\epsilon = +1$.} \label{f1}
\end{figure}
\begin{figure}[t]
\includegraphics[width=6.7cm,height=5.58cm,angle=0]{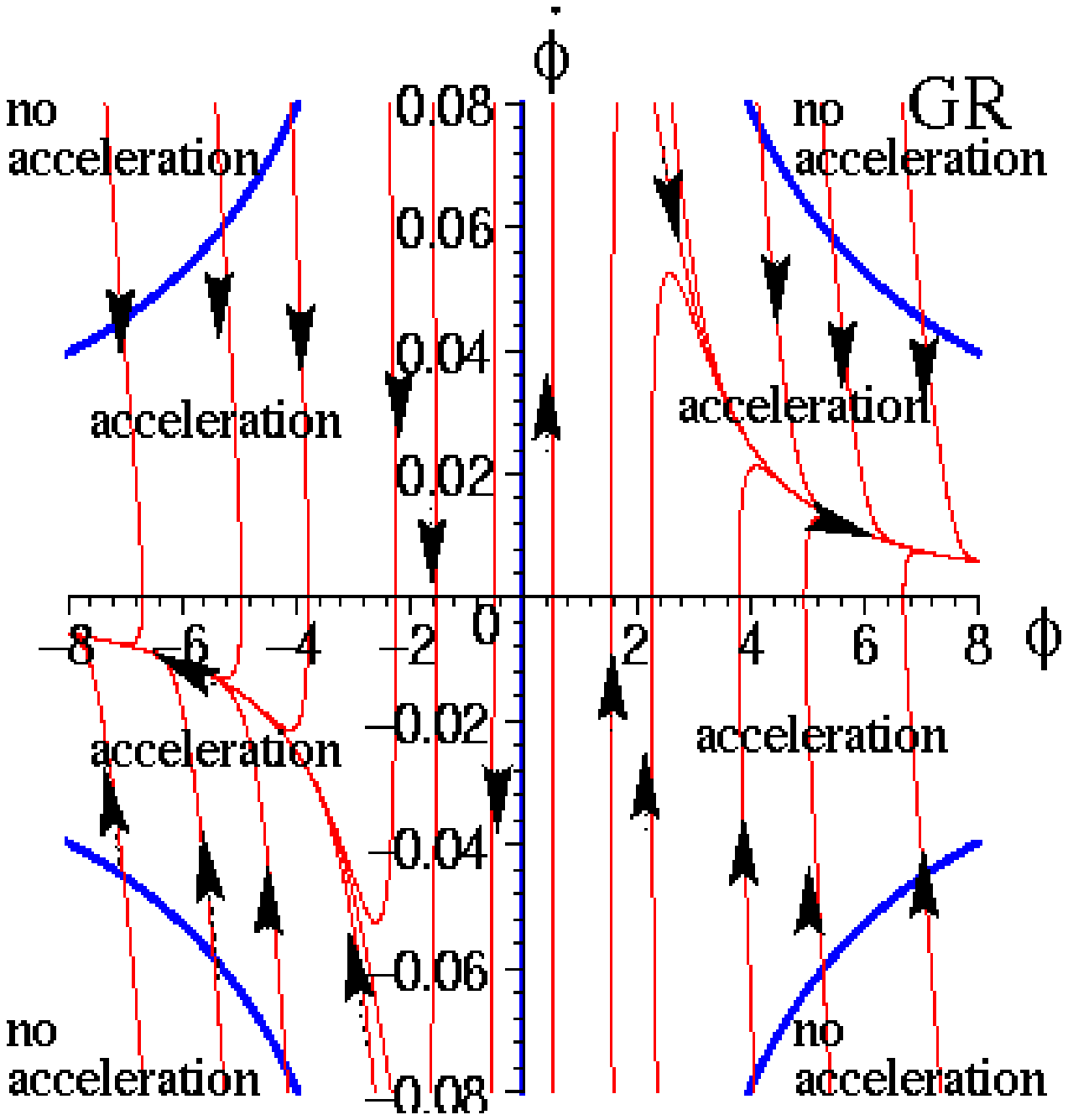}
\;\;\;\;\;\;\;\;\;\;
\includegraphics[width=6.7cm,height=5.58cm,angle=0]{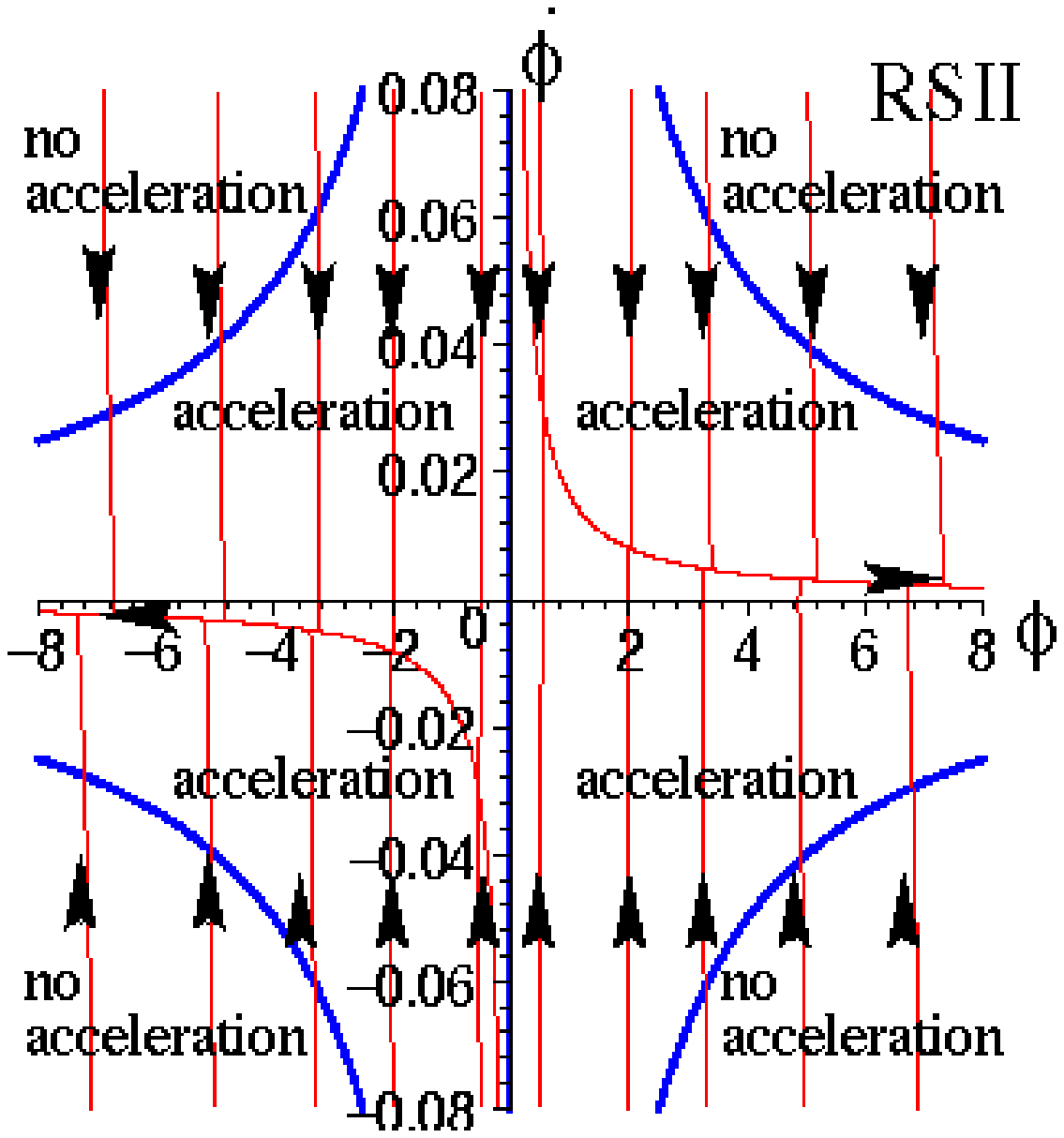} \\ \vspace{0.7cm}
\includegraphics[width=6.7cm,height=5.58cm,angle=0]{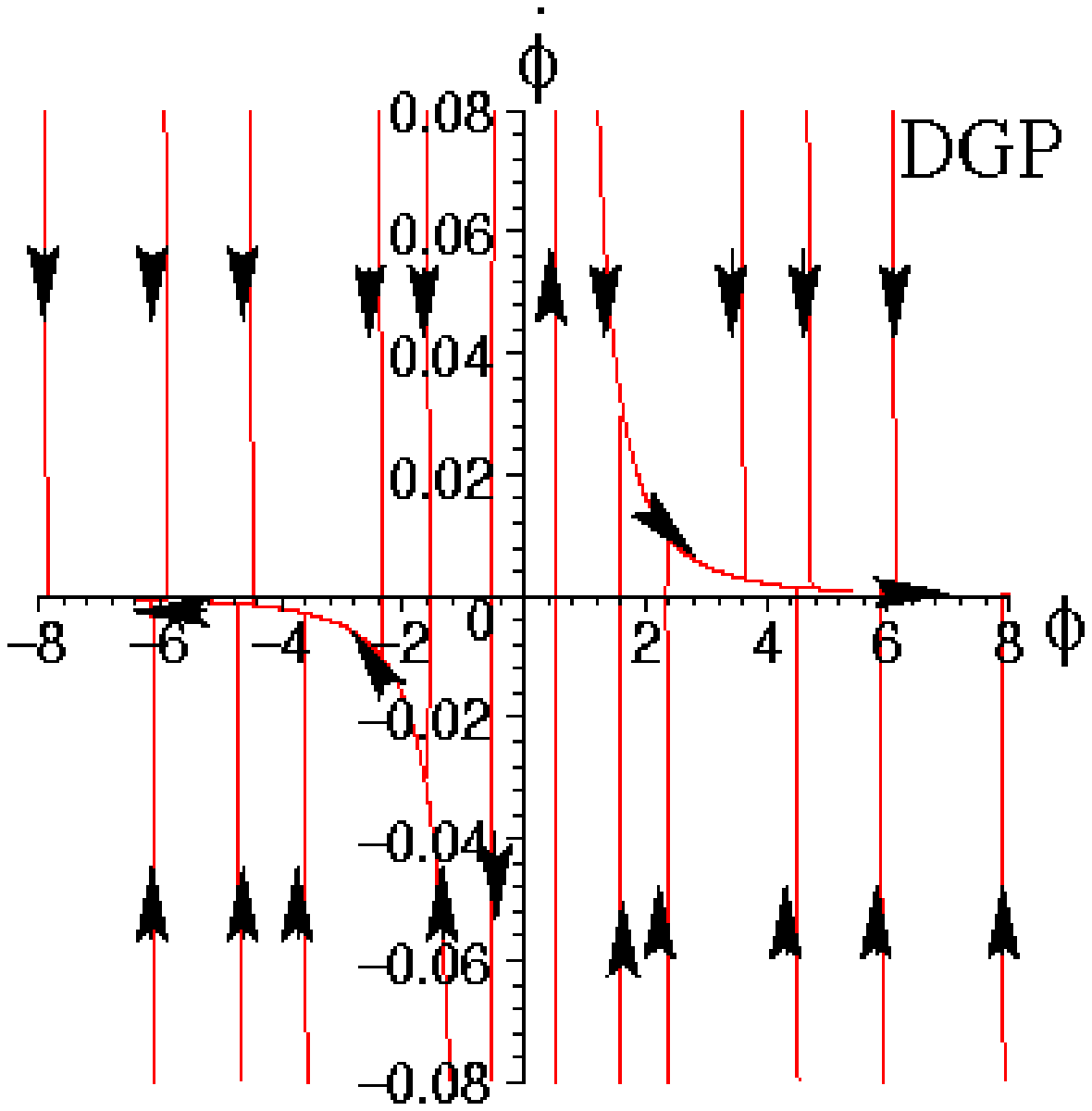}
\caption{The phase space of a scalar field with $V(\phi)=\mu^{n+4}
/ {\phi}^n $, $n=2$,  for the standard cosmology, RSII and DGP
brane cosmologies. The curves marking the acceleration condition
are the solid lines. (The acceleration curve for the DGP case is
beyond the scales on the axes; see Fig. \ref{fextra}.)} \label{f2}
\end{figure}

\section{Braneworld scalar field phase space}
The energy density and pressure of the scalar field are $\rho =
\frac{1}{2}\dot{\phi}^2 + V(\phi)$ and $ p =
\frac{1}{2}\dot{\phi}^2 - V(\phi)$. When the scalar field
dominates the universe, the conservation equation takes the form
of the Klein-Gordon equation
\begin{equation}
\ddot{\phi} + 3H\dot{\phi} = - \frac{dV}{d\phi}
\end{equation}
Inflation requires a scalar field to drive the acceleration. The
standard acceleration condition for a scalar field dominated
universe is $ \dot{\phi}^2 < V(\phi)$. In the RSII model when a
scalar field dominates, the condition can be derived from equation
(\ref{rsac}) to obtain (see Ref. \cite{mwbh})
\begin{equation}
{\dot{\phi}}^2 - V + \frac{{\dot{\phi}}^2 + 2V}{8 \lambda} \big( 5
{\dot{\phi}}^2 - 2V \big) \; < \; 0  \label{scalarcon}
\end{equation}
The third term on the left is the modified effect from the
quadratic density term in the Einstein equation \cite{mwbh, tmm}.
There are some classes of potentials that are too steep to give
inflation in standard cosmology. However in the RSII scenario with
the presence of the additional quadratic term in the Friedmann
equation, inflation can occur and ends with a kinetic regime when
the field has stiff equation of state \cite{cll}.

In the case of the DGP braneworld, since equation (\ref{dgpac})
reduces to standard cosmology at high energy, it gives the usual
acceleration condition of the standard cosmology in the early
universe. Considering a scalar field dominated late universe and
using the approximate condition (\ref{ppp1}), when $\epsilon = +1$
we obtain
\begin{equation}
{\dot{\phi}}^2 - V - \frac{3}{2 r_{c}^2 \kappa^2} \; < \; 0
\label{s1}
\end{equation}
The scalar field needs less potential energy than in the standard
case in order to accelerate the expansion. Acceleration can happen
even if $\dot{\phi}^2$ is equal to or larger than $V$, as long as
it is smaller than $V + 3/2 r_{c}^2 \kappa^2$. However at very
late times, when the field is moving very slowly, we can use the
slow-roll approximation $\ddot{\phi}\sim 0 $ and we can
approximate that ${\dot{\phi}}^2 \ll V + 3/2 r_{c}^2 \kappa^2 $.
Applying these conditions to the DGP Friedmann equation
(\ref{eppos1}) and the Klein-Gordon equation, we obtain
\begin{equation}
\dot{\phi}^2 \; \simeq \; \frac{{V'}^2}{6\kappa^2 V + 9/r_{c}^2}
\label{aj1}
\end{equation}

This equation is the late time attractor trajectory in phase
space. In the case $\epsilon = -1$, using the condition
(\ref{ppp2}) we obtain
\begin{equation}
5 {\dot{\phi}}^2 - 2V  \; < \; 0 \label{dgpepnegcon}
\end{equation}
Applying the DGP Friedmann equation (\ref{epneg1}) and the
Klein-Gordon equation at very late time, with similar assumptions
as above, a late time phase space attractor trajectory can be
found from
\begin{equation}
\dot{\phi}  \; \simeq \; - \frac{1}{r_c \kappa^2} \frac{V'}{V}
\label{aj2}
\end{equation}
Now we consider the phase space for some particular types of
potentials $V(\phi)$. Here we will consider mainly the DGP
braneworld model but illustration of the phase-space in standard,
RSII and DGP models will be presented for comparison. In our
numerical plots, we  include the curve showing the acceleration
condition. The curve divides  acceleration from non-acceleration
regions.

\subsection{Inverse-power potential}
The potential
\begin{equation}
 V(\phi) = \frac{\mu^{n+4}}{{\phi}^n}
\end{equation}
  is one candidate for a quintessential
potential. This potential was considered in some supersymmetry
breaking models \cite{qsusy}. The potential can give tracking
behavior in which the scalar field energy density evolves along
with the density of matter and radiation \cite{zws}. The ratio of
scalar field density and total density of matter and radiation
increases slowly as $\rho_{\phi}/(\rho_m + \rho_r) \propto
t^{4/(n+2)}$ \cite{s}. For early universe inflation, this
potential is too steep for standard inflation. However the RSII
braneworld scenario can give inflation at high energy but with
large magnitude relic gravitational waves \cite{ss}. Solutions of
the scalar field dominated era and radiation dominated era of RSII
brane cosmology have been found \cite{mmy}. In this subsection we
consider phase spaces of scalar field dominated universe and their
acceleration conditions in three cases: early universe RSII brane
cosmology, low energy standard cosmology limit of RSII brane
cosmology, and DGP brane cosmology. Since observations constrain
$n \lesssim 2$ \cite{Balbi:2000kj}, we choose $n=1$ and $2$ in our
numerical plots.

\subsubsection{High energy regime of RSII brane cosmology}
During the high energy regime, $\rho \gg \lambda$, the quadratic
term in the RSII Friedmann equation is dominant. From the
acceleration condition in equation (\ref{scalarcon}), dropping the
first two terms at high energy and using a slow-roll approximation
($\ddot{\phi}\sim 0$), one can get the attractor trajectory
\cite{mmy}
\begin{equation}
\phi \: \dot{\phi} = \frac{n}{3\kappa}\sqrt{6 \lambda}
\label{at1phi}
\end{equation}

For $n \leq 2$ inflation can happen. For $n=2$ the kinetic term
and potential terms are balanced and this yields power law
expansion. For $n > 2$ the kinetic term is dominant (see
\cite{Maeda:2000mf, mmy}). The attractor trajectory equation
(\ref{at1phi}) matches the RSII phase plot of Figs. \ref{f1} and
\ref{f2} at late times.

\subsubsection{Low energy standard cosmology limit of RSII brane
cosmology}  In the late universe, the RSII Friedmann equation
approaches the standard cosmology limit. When a quintessential
scalar field dominates the universe at late times, the Friedmann
equation (with small value of $\dot{\phi} $) is
\begin{equation}
H = \frac{\kappa}{\sqrt{3}}\left(
\frac{\mu^{n+4}}{{\phi}^n}\right)^{1/2}
\end{equation}
Together with the Klein-Gordon equation in the slow-roll regime,
we find that the attractor curve is described by
\begin{equation}
\dot{\phi}\:\phi^{\frac{n}{2} +1} = \frac{n }{\sqrt{3}\:\kappa}\:
\mu ^{\frac{n}{2} +2} \label{at2phi}
\end{equation}
The attractor trajectory agrees well with the numerical plots at
late times as seen in the standard cosmology case in Figs.
\ref{f1} and \ref{f2}.

\begin{figure}[t]
\includegraphics[width=6cm,height=5cm,angle=0]{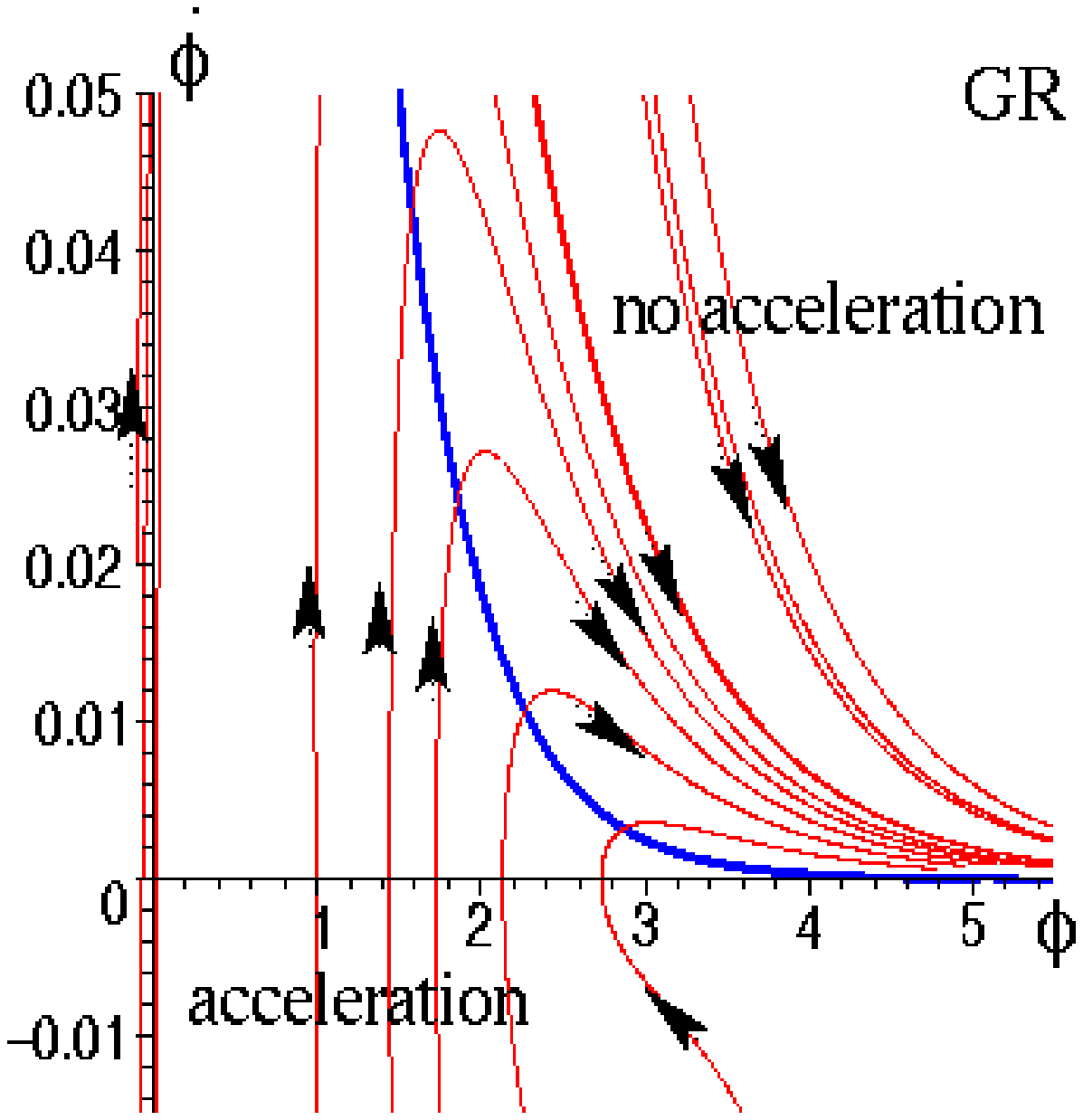}
\;\;\;\;\;\;\;\;\;\;
\includegraphics[width=6cm,height=5cm,angle=0]{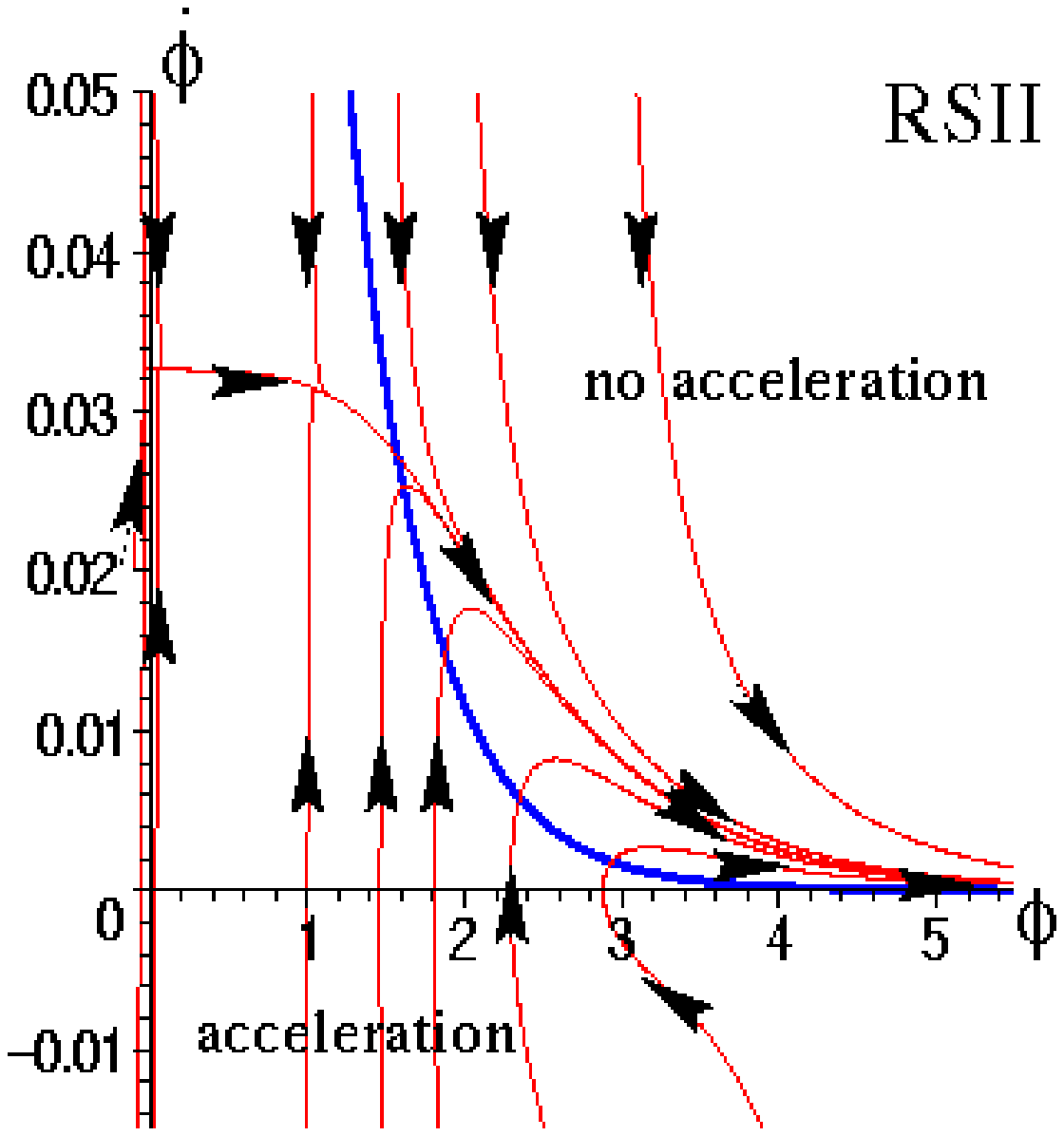}
\\
\vspace{0.7cm}
\includegraphics[width=6cm,height=5cm,angle=0]{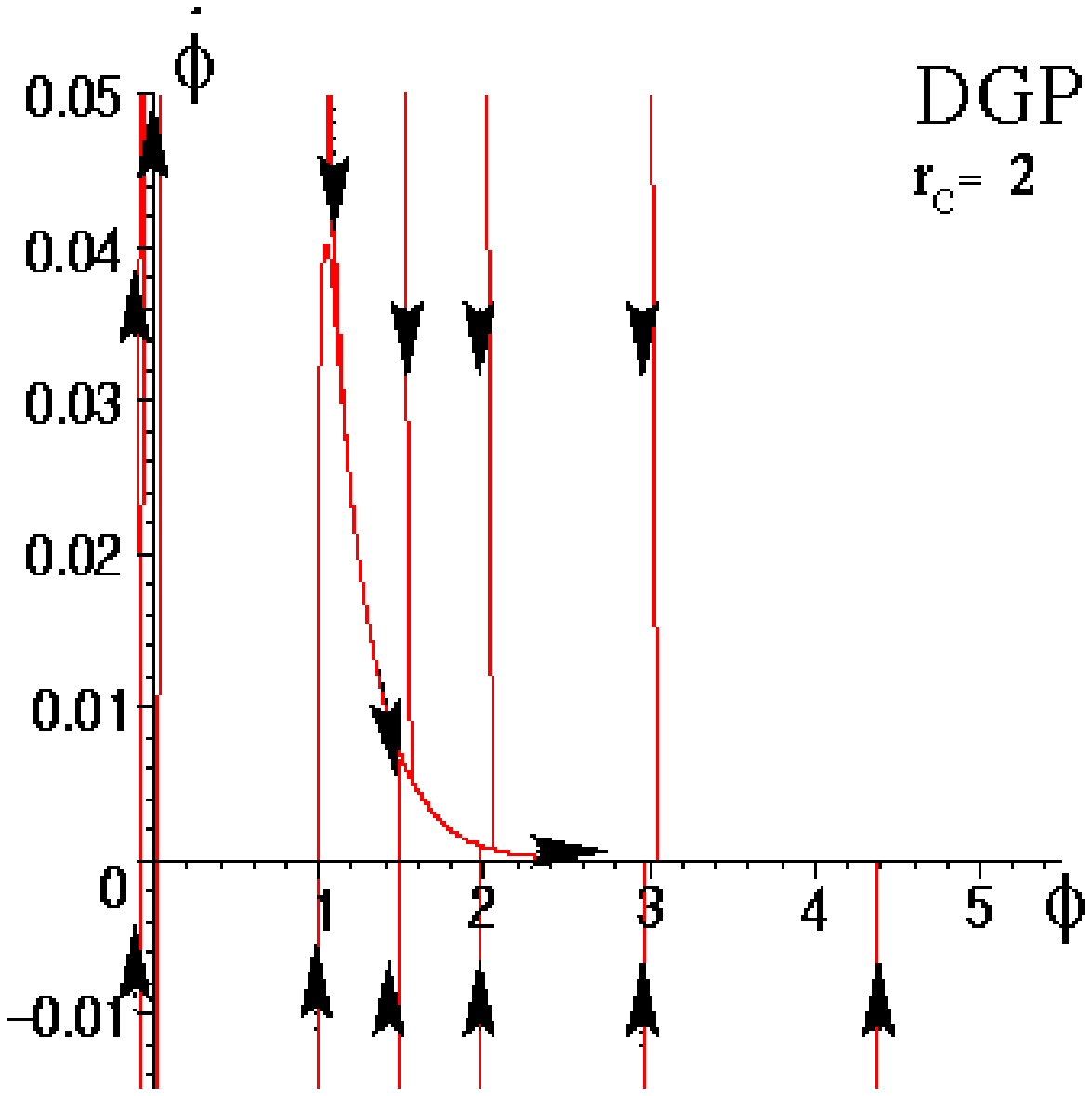}
\caption{The phase space of a scalar field with exponential
potential, equation (\ref{exponent}), for the standard cosmology,
RSII and DGP brane cosmologies. The solid lines are the
acceleration condition curves. (The acceleration curve for the DGP
case is beyond the scales on the axes; see Fig. \ref{fextra}.) We
use $ \kappa=1 $, $p=1/8$, $\lambda = 10^{-4}$, $r_c = 2$, and $
V_0 = 1 $.} \label{f3}
\end{figure}
\begin{figure}[t]
\begin{center}
\includegraphics[width=8cm,height=7cm,angle=0]{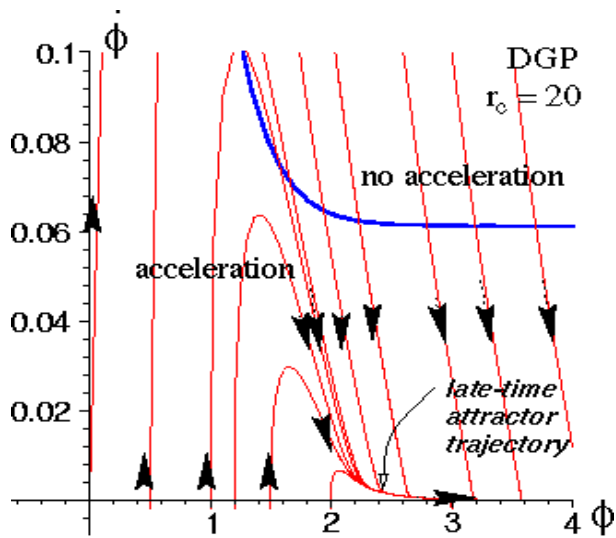}
\end{center}
\caption{The phase space of a scalar field with exponential
potential, equation (\ref{exponent}), in a DGP brane universe with
$r_c = 20$. The crossover scale term dominates the potential term
at $\dot{\phi} \simeq 0.061$. The solid line is the acceleration
condition curve.} \label{fextra}
\end{figure}

\subsubsection{DGP brane cosmology}
DGP cosmology recovers standard cosmology in the early universe
when $\rho \gg 3/4r_{c}^2 \kappa^2$. However when the
extra-dimension effects become significant in the late universe,
i.e. $\rho \leq 3/4r_{c}^2 \kappa^2$, the modification of its
dynamics is crucial. We here consider the low-energy late universe
with scalar field domination. Consider the case $\epsilon = +1$.
Using equation (\ref{aj1}), the slow-roll attractor trajectory at
late times is
\begin{equation}
{\dot{\phi}}^2   \; = \; \frac{n^2 \mu^{2n+8} \phi^{-2n-2}}{6
\kappa^2 \left[  \mu^{n+4} \phi^{-n} + 3/2 r_{c}^2 \kappa^2
\right]} \label{DGPphiminusn}
\end{equation}
At small $\dot{\phi}$, equation (\ref{DGPphiminusn}) fits the
numerical plots in both $n=1$ and $2$ cases, as seen in Figs.
\ref{f1} and \ref{f2}.

In the case $\epsilon = -1$, the attractor trajectory at late time
is
\begin{equation}
\phi \: \dot{\phi} \;=\; \frac{n}{r_c \kappa^2}
\end{equation}
and the solution is $ \phi = \left[2n(t-\tau)/\kappa^2 r_{c}
\right]^{1/2} $. Indeed in the case $\epsilon = -1$, the equation
(\ref{epneg1}) has a similar form to that of RSII at high
energies. Therefore, all subsequent results looks similar to those
of the RSII case which has been found before (see \cite{mmy}). For
example in our $\epsilon = -1$ DGP case, the slow-roll condition
can be applied only when $n<2$ since $\ddot{\phi} \:(\sim
t^{-3/2}) \ll V'\: (\sim t^{-(n+1)/2} )$ and $\dot{\phi}^2 \:(\sim
t^{-1}) \ll V \:(\sim t^{-n/2})$.

\begin{figure}[t]
\begin{center}
\includegraphics[width=6cm,height=5cm,angle=0]{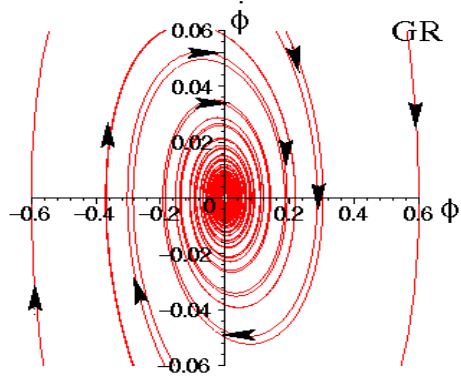}
\end{center}
 \caption{{The phase space of a scalar field with $V(\phi)=V_0
\left[\cosh(\alpha \phi/M_4)-1\right]$ in standard cosmology. The
numerical values are $\kappa=1$, $\alpha =3$ and $V_0 = 5 \times
10^{-3}$.}} \label{f4}
\end{figure}
\begin{figure}[t]
\begin{center}
\includegraphics[width=6cm,height=5cm,angle=0]{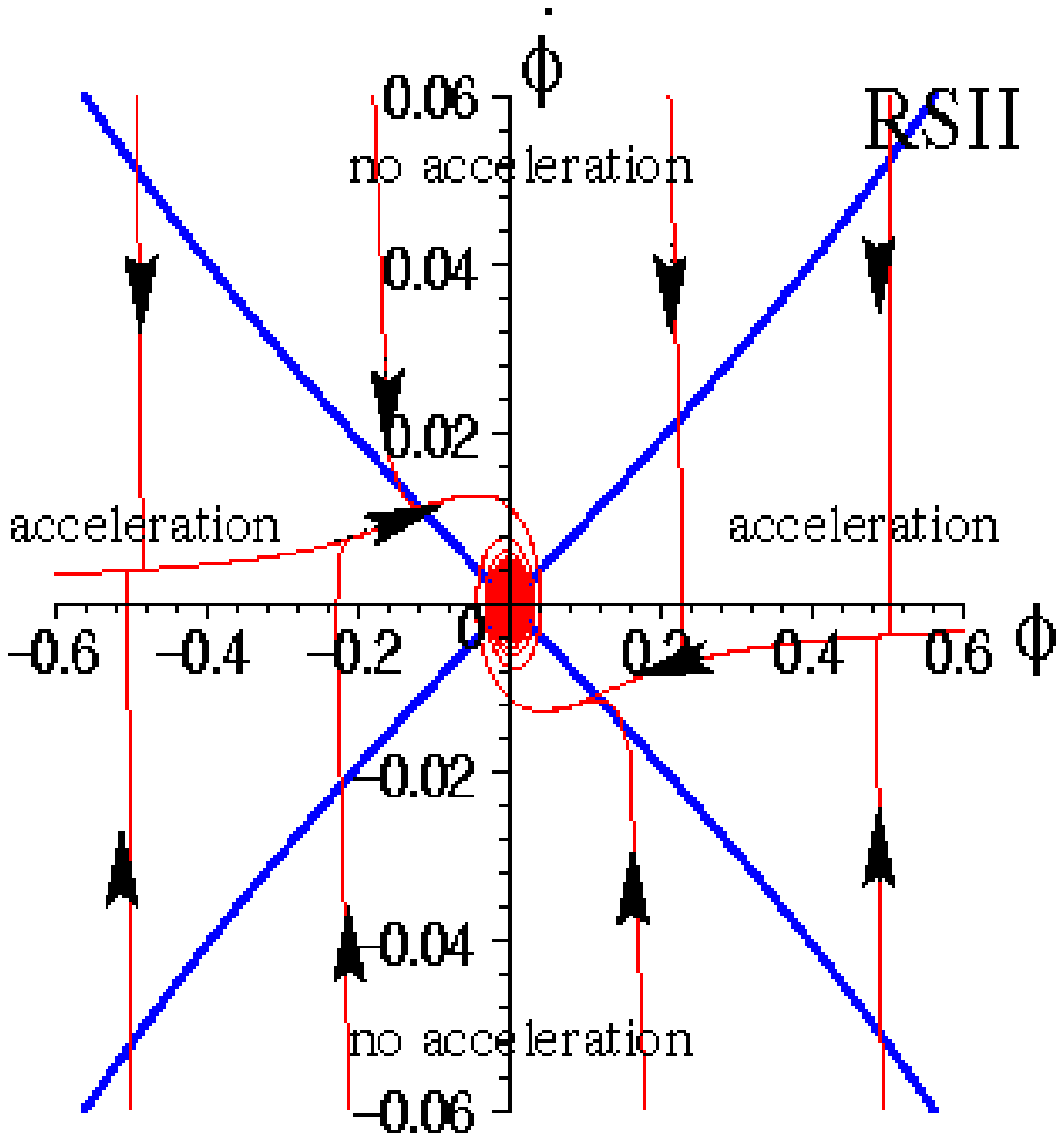}\;\;
\;\;\;\;\;\;\;\;\;\;
\includegraphics[width=6cm,height=5cm,angle=0]{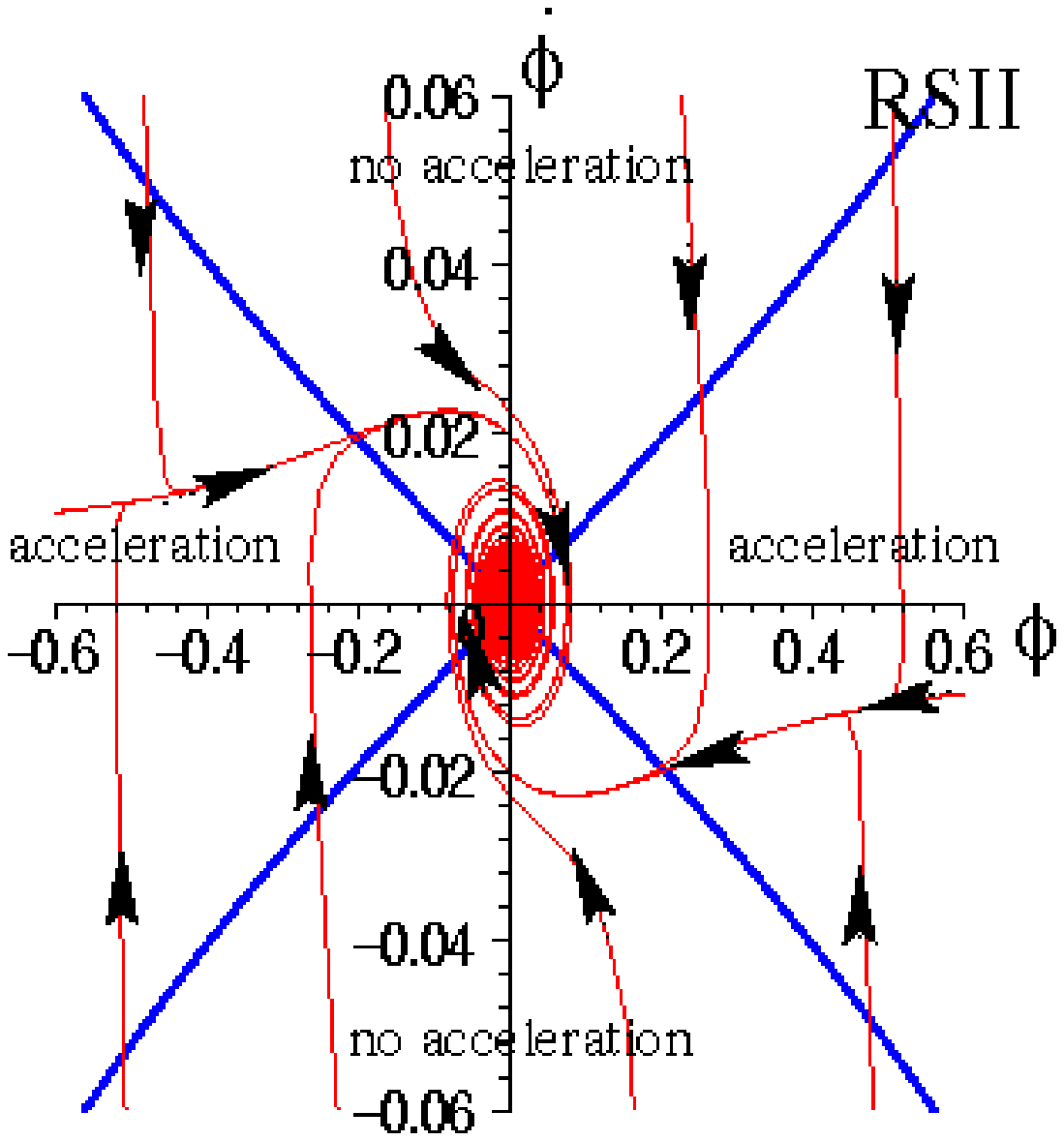}\;\;
\\ \vspace{0.7cm}
\includegraphics[width=6cm,height=5cm,angle=0]{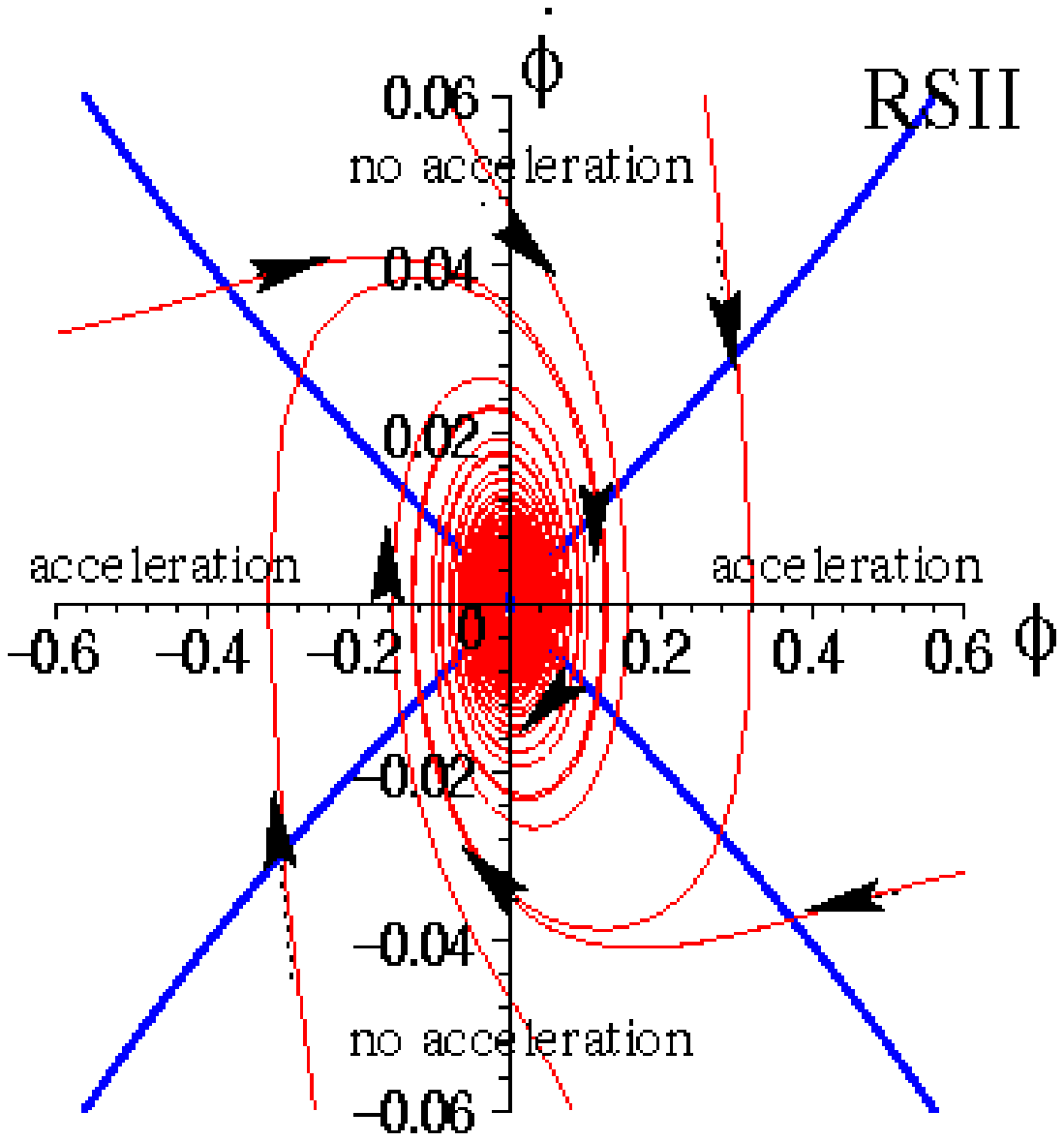}\;\;
\;\;\;\;\;\;\;\;\;\;
\includegraphics[width=6cm,height=5cm,angle=0]{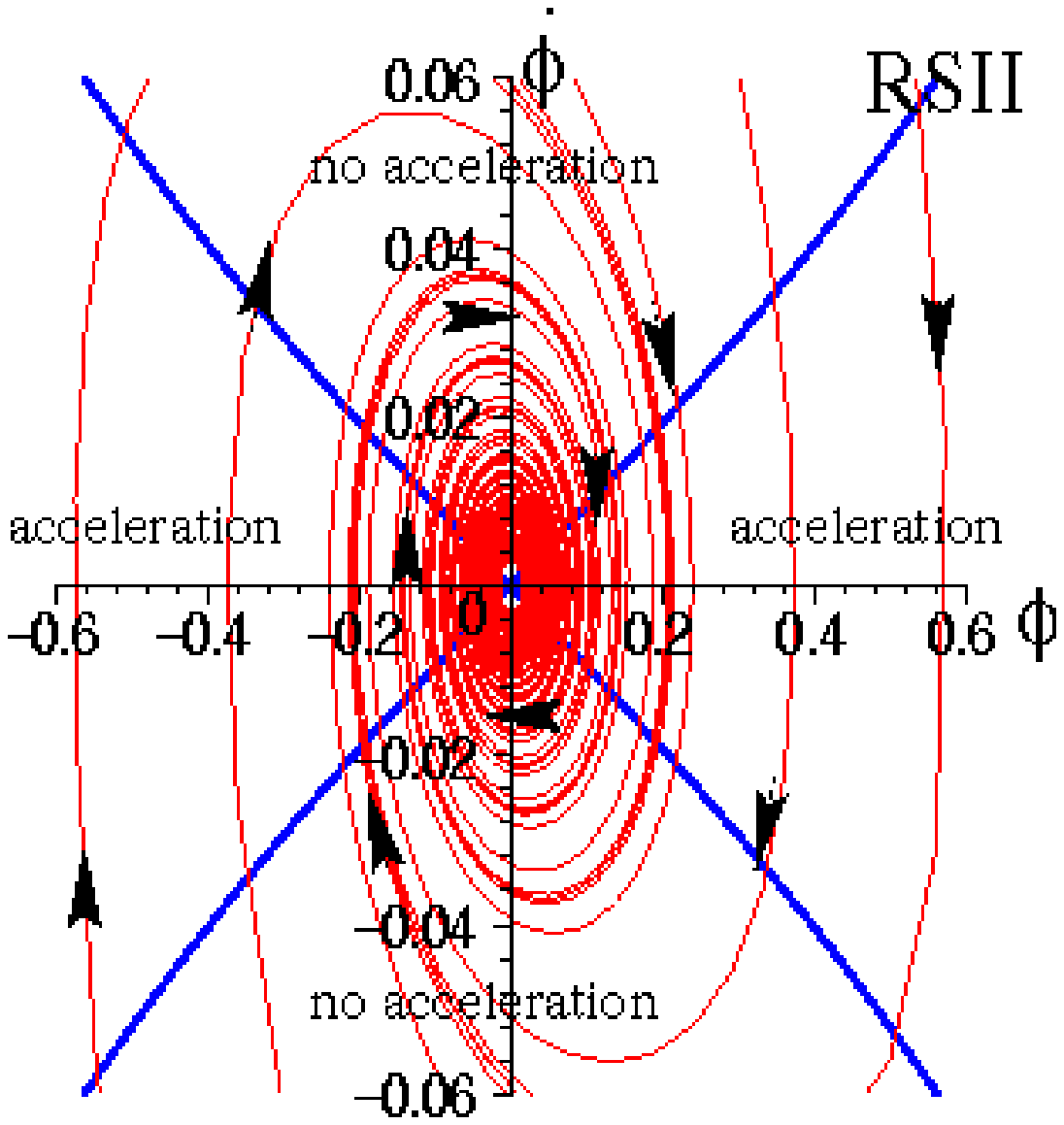}
\end{center}
 \caption{The phase space of a scalar field with
$V(\phi)=V_0 \left[\cosh(\alpha \phi/M_4)-1\right]$ in RSII brane
cosmology, with $\lambda = 10^{-6}, 10^{-5}, 10^{-4}$ and
$10^{-3}$. The solid lines are the acceleration condition.}
\label{f5}
\end{figure}
\begin{figure}[t]
\begin{center}
\includegraphics[width=6cm,height=5cm,angle=0]{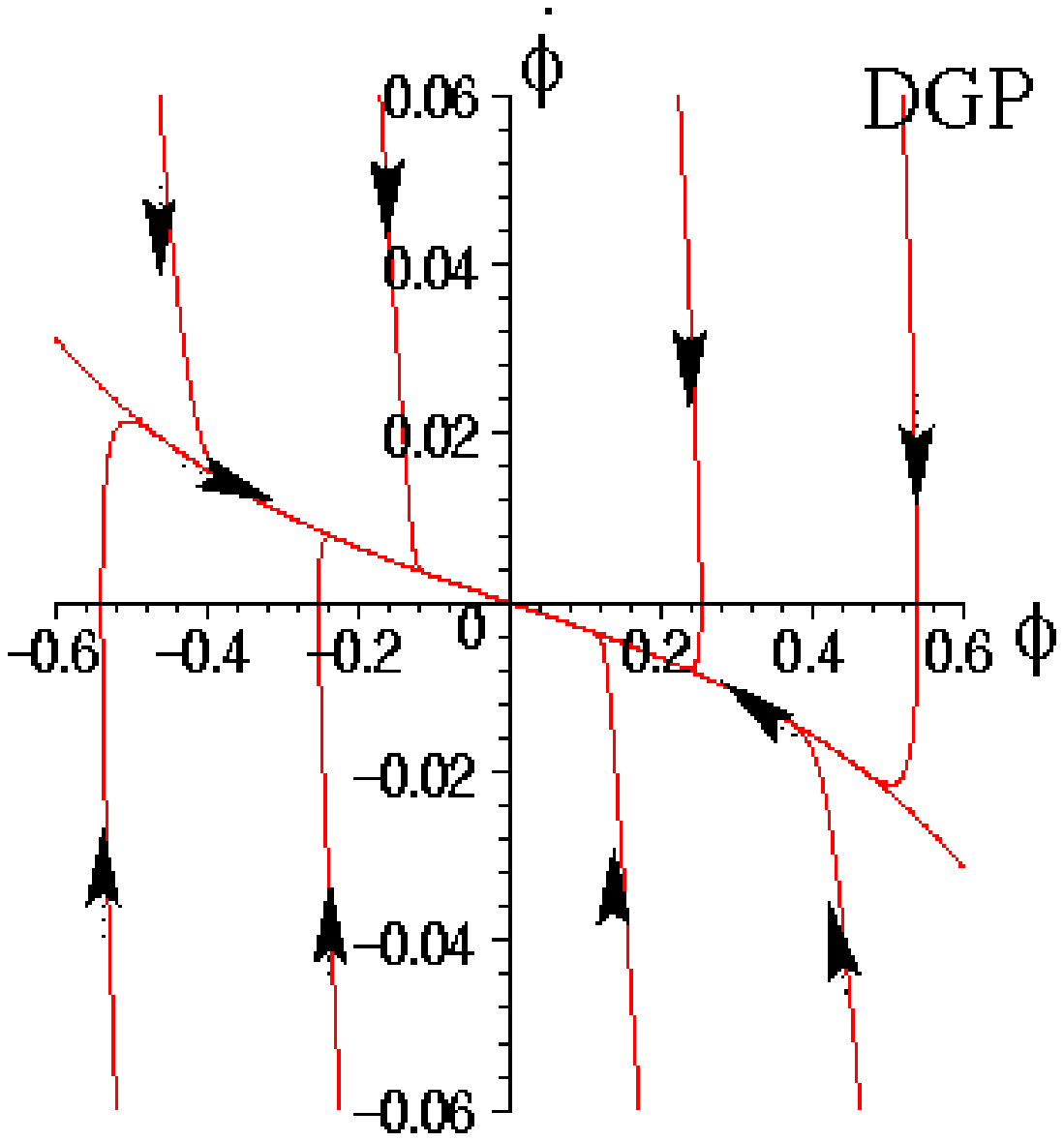}
\;\;\;\;\;\;\;\;\;\;
\includegraphics[width=6cm,height=5cm,angle=0]{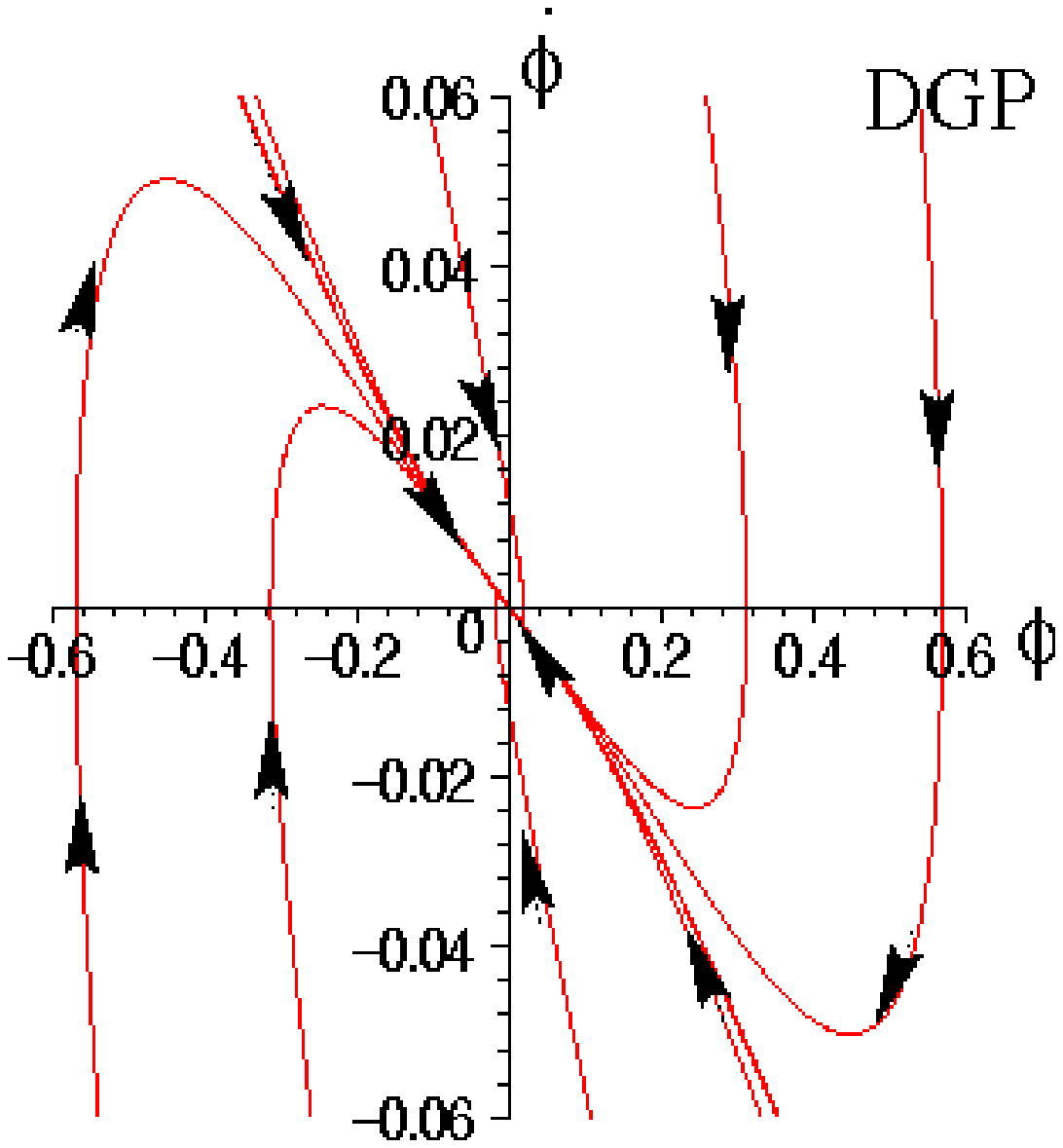}
\\ \vspace{0.7cm}
\includegraphics[width=6cm,height=5cm,angle=0]{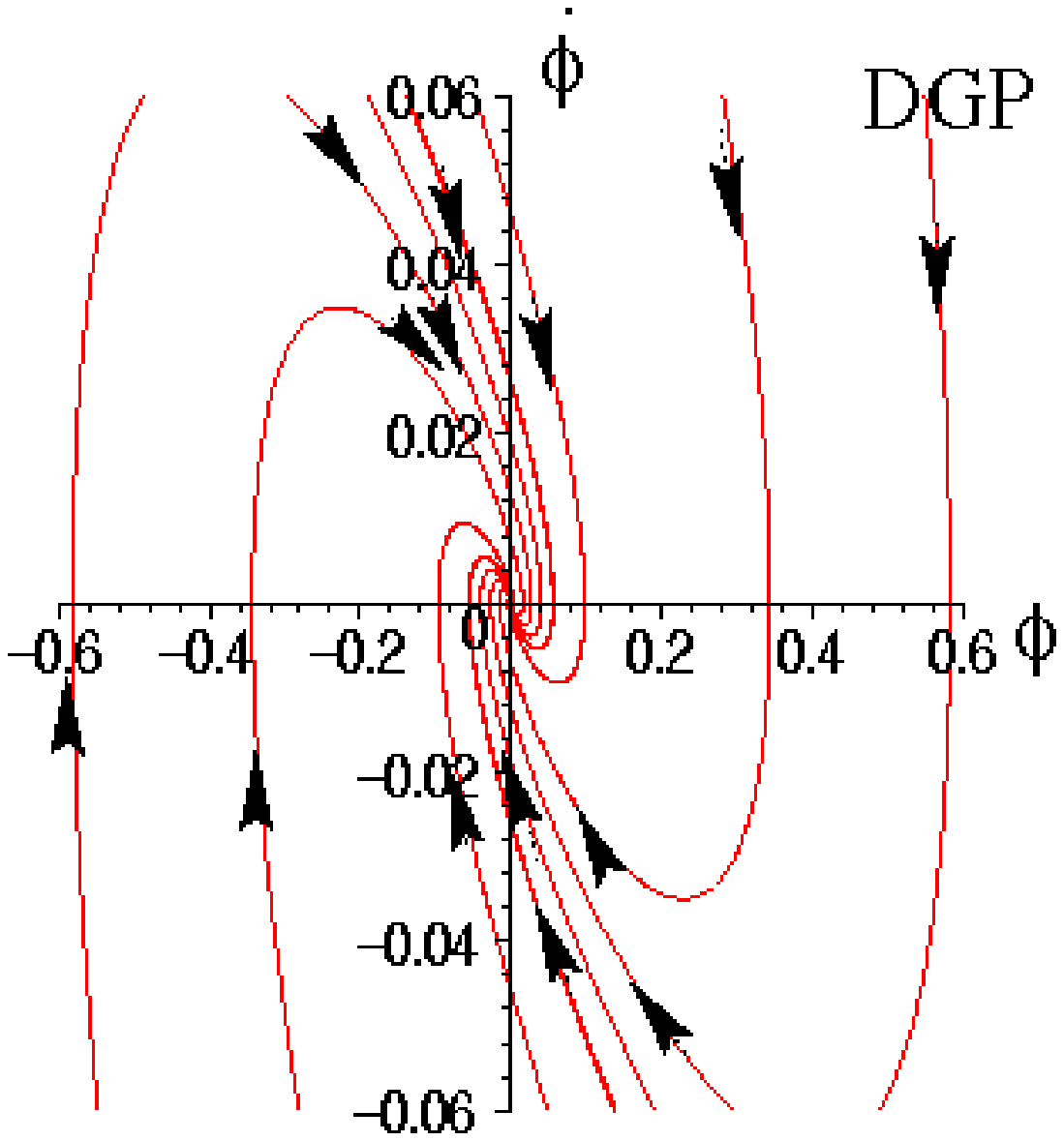}
\;\;\;\;\;\;\;\;\;\;
\includegraphics[width=6cm,height=5cm,angle=0]{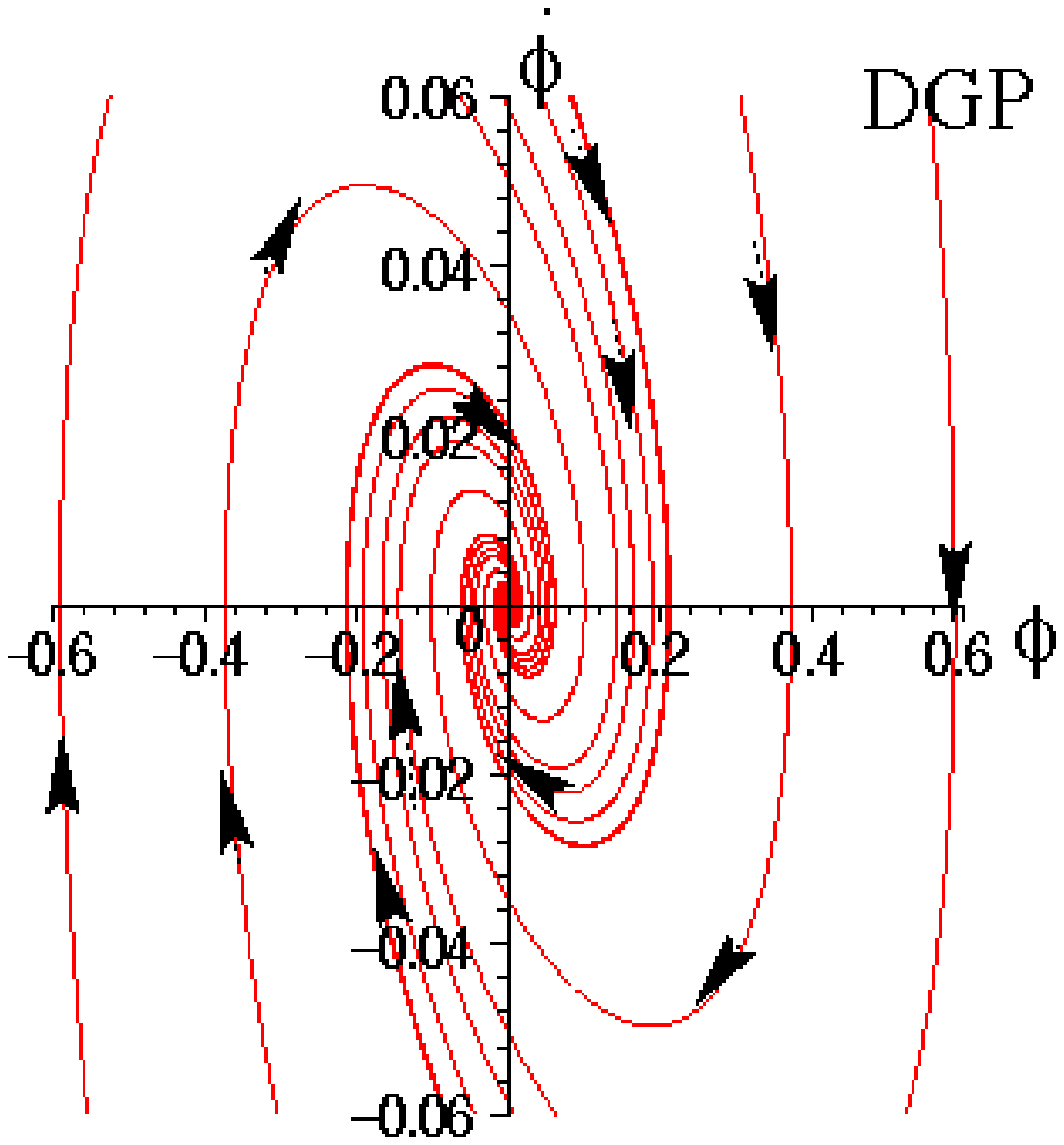}
\end{center} \caption{The phase space of a scalar field with
$V(\phi)=V_0 \left[\cosh(\alpha \phi/M_4)-1\right]$ in DGP brane
cosmology, with $r_{c} = 2, 6, 10$ and $20$.} \label{f6}
\end{figure}

\subsection{Exponential potential}
A scalar field with exponential potential
\begin{equation}
V(\phi) = V_{0} \exp\left(-\sqrt{\frac{2}{p}}
\frac{\phi}{M_4}\right) \label{exponent}
\end{equation}
 is known to yield power-law inflation in standard cosmology, i.e. $a
= a_0(t/t_0)^p$ \cite{lm}. The slow-roll parameters in this case
are $ \varepsilon = \eta/2 = 1/p $ and the inflation condition
requires that $ p>1 $ \cite{liddle89}. The exponential potential
that drives inflation could be the same potential that is
responsible for quintessence and cold dark matter unless the
quintessential and cold dark matter potentials are too steep
($p<1$) for inflation to happen \cite{rp}. As quintessence, an
exponential potential can yield tracking behavior. However it
evolves with constancy of the ratio between background matter
density and scalar field density, so it is not able to yield a
quintessence dominated universe at late time. Some modification of
this model for standard cosmology has been proposed, e.g. the
potential $V(\phi) = V_0 \left[\cosh(\alpha \phi/M_4)-1\right]^q$
\cite{s,sw}.

In the RSII braneworld, a steep potential, i.e. $p<1$, is able to
drive inflation. The quadratic term in the RSII Friedmann equation
is dominant at high energy, contributing to more friction in the
Klein-Gordon equation. As a result, slow-roll inflation can be
obtained. After the quadratic term stops dominating, the field
moves faster and enters a kinetic regime where the inflaton field
can redshift faster than the produced particles and radiation,
i.e. $\rho_{\phi}\propto a^{-6}$. This results in a natural exit
and reheating by gravitational particle production \cite{cll}.

   In the DGP brane universe, acceleration of the universe is an
effect from both quintessence and existence of the
extra-dimension. Using equation (\ref{aj1}), the slow-roll
attractor trajectory in phase space is
\begin{equation}
\dot{\phi}^2 \;=\; \frac{\alpha^2 V_0 e^{-\alpha \phi}} { 6
\kappa^2    + 9 e^{\alpha \phi}/V_0 r_{c}^2 } \label{exptra}
\end{equation}
where $\alpha = \sqrt{2/p}\: M_4^{-1}$. The standard general
relativity limit of equation (\ref{exptra}) is recovered when
$r_{c} \rightarrow \infty $, and yields the standard cosmology
solution $\phi(t) \sim \ln(t) $. The full solution of equation
(\ref{exptra}) is
\begin{equation}
t-\tau \;=\; \frac{\sqrt{A e^{\alpha \phi} +B e^{2\alpha\phi}}}
{\alpha} + \frac{A}{2 \alpha \sqrt{B}} \ln \left\{
\frac{8}{\alpha^2}\left[  \frac{A}{2}+ B e^{\alpha \phi}
\:+\:\sqrt{B} \sqrt{  A e^{\alpha \phi} + B e^{2\alpha \phi}}
\right]\right\}
\end{equation}
where $A = 6 \kappa^2/\alpha^2 V_{0}$ and $B = (3/ \alpha V_{0}
r_c )^2 $. The solution is found when $B$ is nonzero.

Numerical plots of the standard, RSII and DGP ($\epsilon = +1$)
braneworld phase spaces are illustrated in Fig. \ref{f3}. In this
figure, it can be seen that the RSII phase space is similar to the
standard cosmology phase space at late time. In RSII case of Fig.
\ref{f3}, if we start from some suitable initial conditions, the
field moves to the attractor trajectory in the acceleration phase
and gives slow-roll inflation. After that, it enters a
non-acceleration phase where the attractor trajectory becomes too
steep for inflation and the field is in a kinetic regime. All
regions of the DGP case of Fig. \ref{f3} are in the acceleration
phase (the non-accelerating region is off the scale of the plot).

In Fig. \ref{fextra} where $r_c = 20$, the DGP acceleration curve
is shown. Due to the high $r_c$ value in this figure, the phase
space looks similar to the standard case at early time, but it
approaches the DGP attractor trajectory at late time. Most of the
area of the figure is in acceleration phase. Moving from any
initial conditions, the field will eventually be in acceleration
phase.

\begin{figure}[t]
\begin{center}
\includegraphics[width=6cm,height=5cm,angle=0]{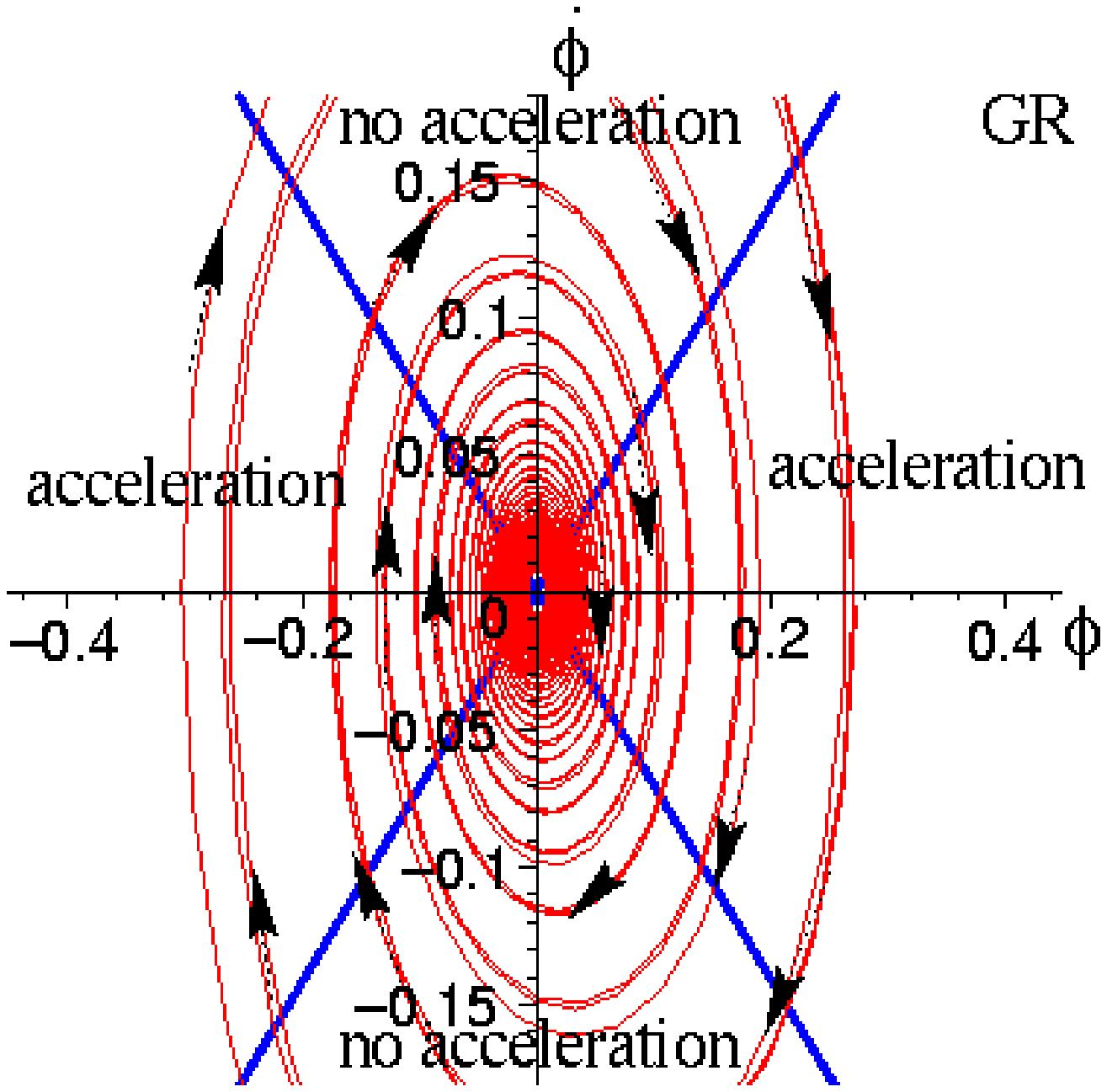}
\;\;\;\;\;\;\;\;\;\;
\includegraphics[width=6cm,height=5cm,angle=0]{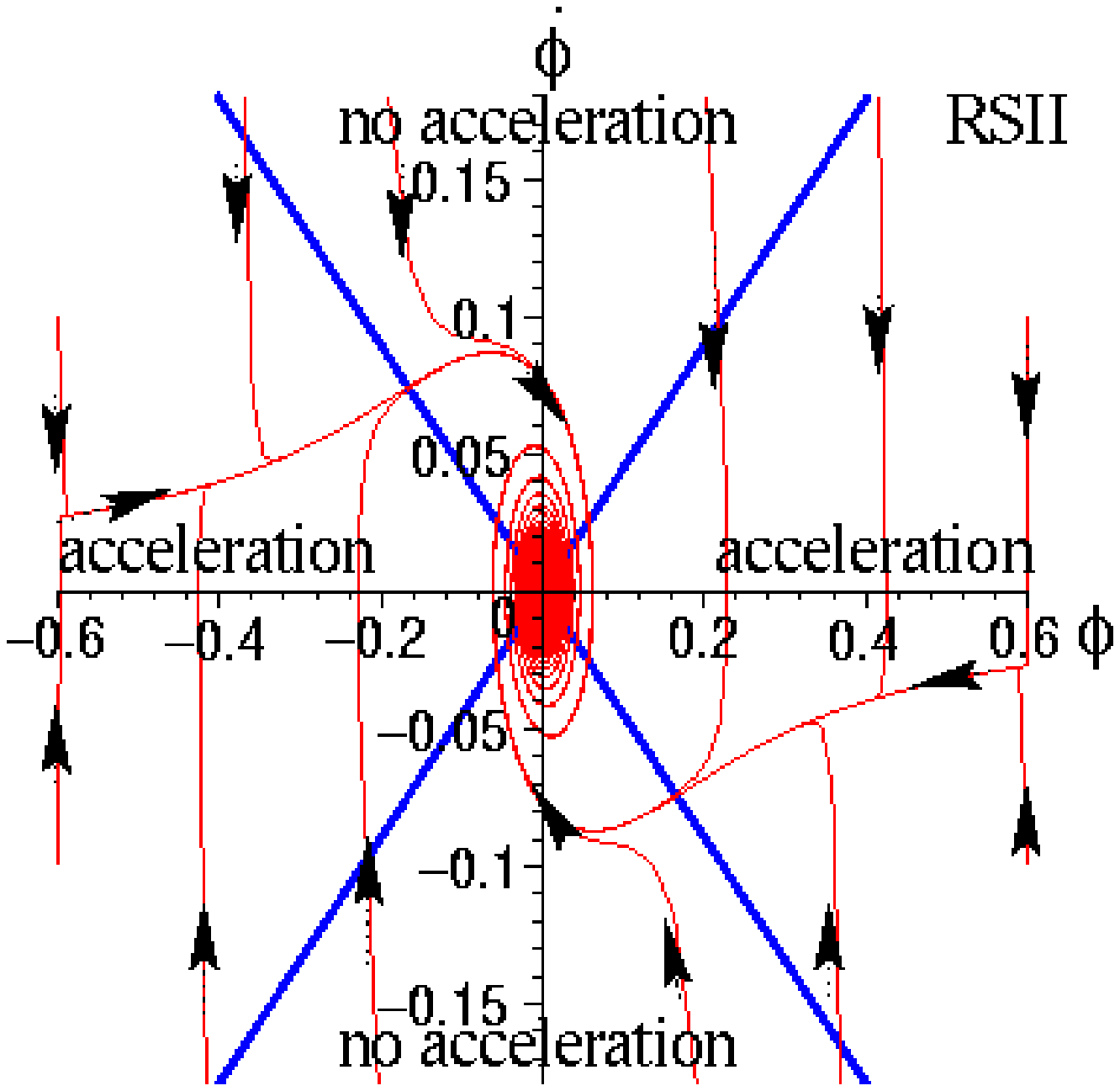}
\\ \vspace{0.7cm}
\includegraphics[width=6cm,height=5cm,angle=0]{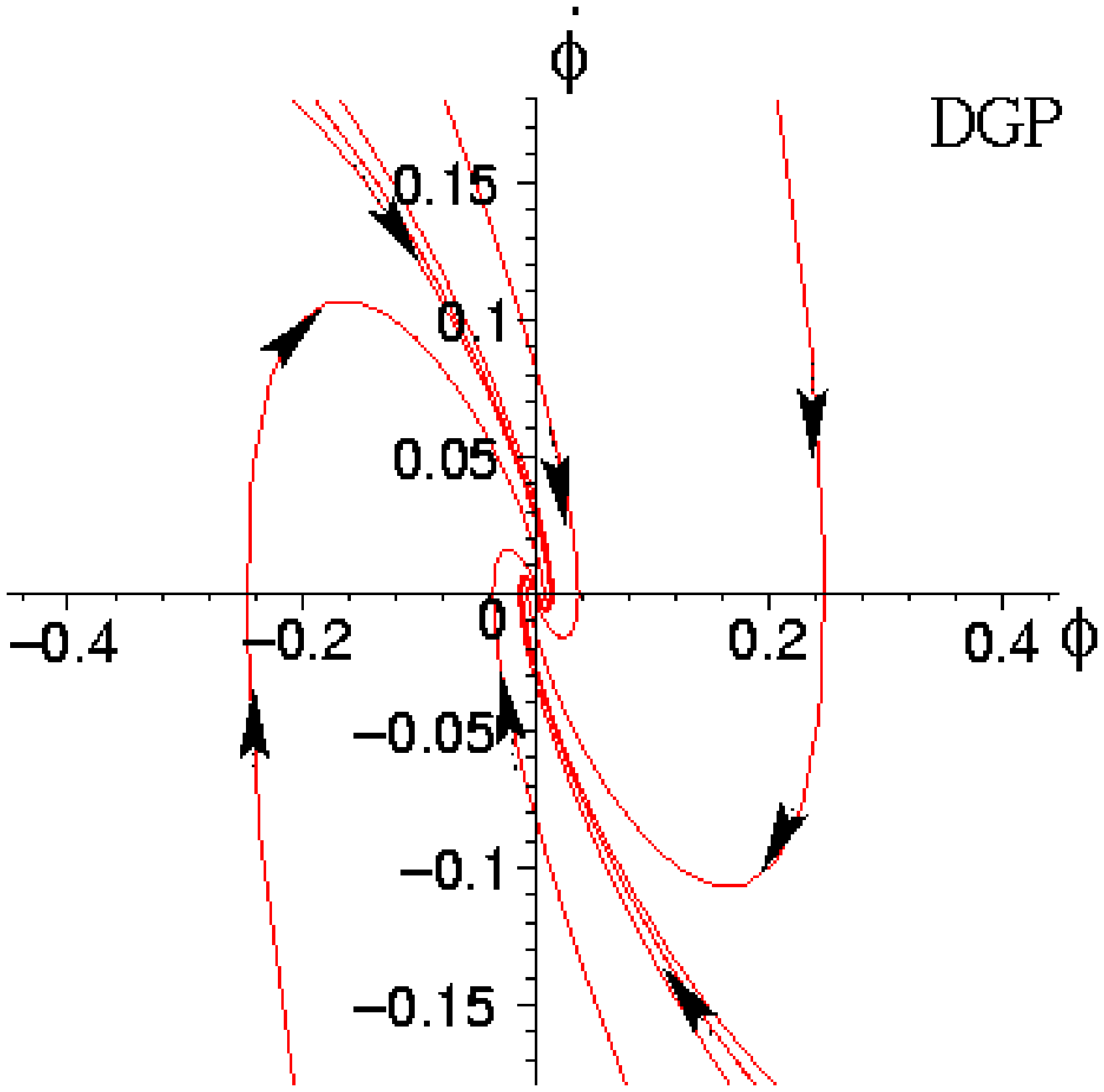}
\end{center} \caption{The phase space of a scalar field with $V(\phi)=\frac{1}{2}m^2 \phi^2 $
for the standard cosmology, RSII and DGP brane cosmologies. The
solid lines are the acceleration condition. We set $\epsilon =
+1$, $ \kappa = m = \sigma =1 $, $\lambda = 10^{-4}$, $r_c = 2$.}
\label{f7}
\end{figure}
\begin{figure}[t]
\begin{center}
\includegraphics[width=6cm,height=5cm,angle=0]{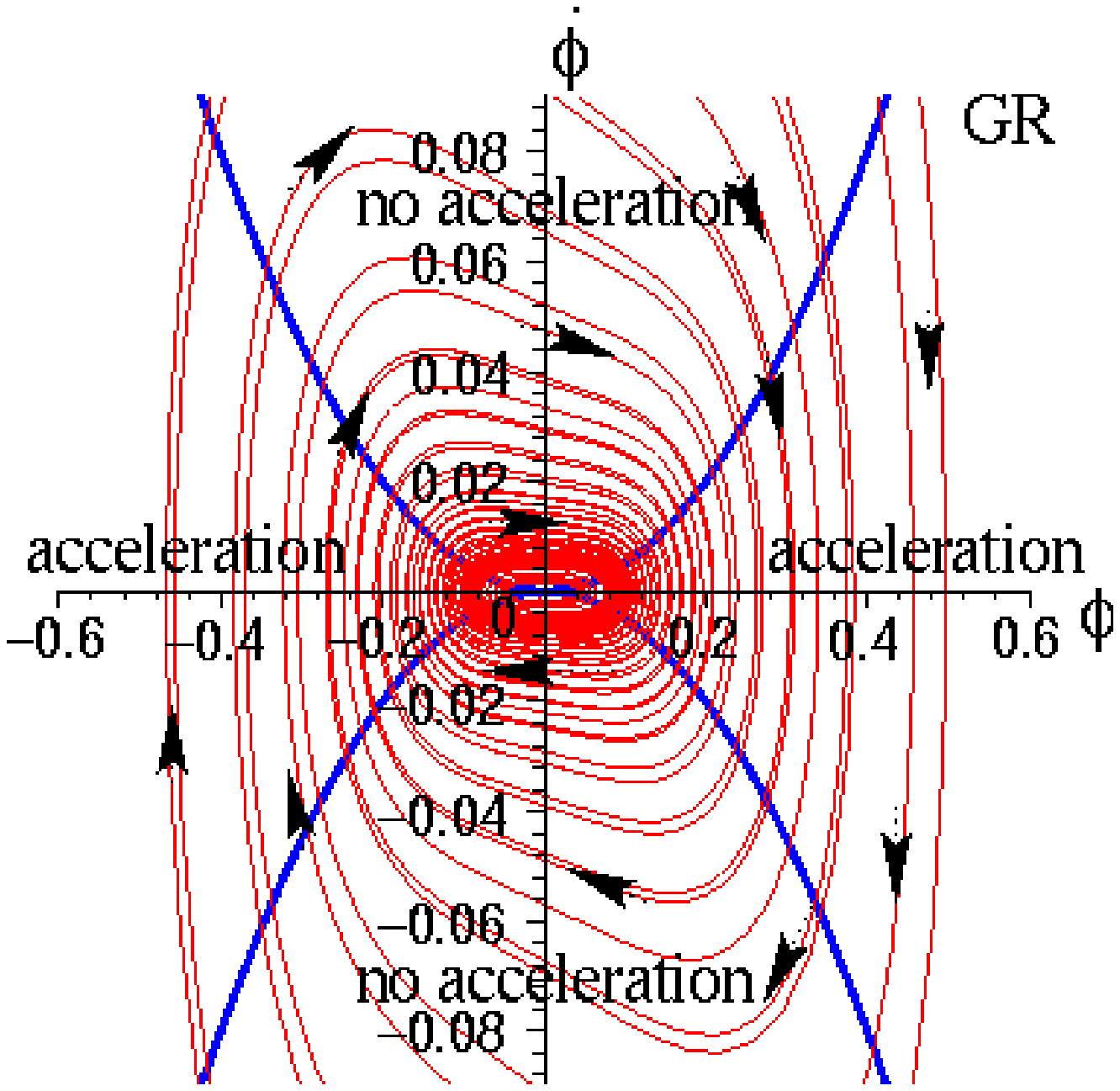} \;\;\;\;\;\;\;\;\;\;
\includegraphics[width=6cm,height=5cm,angle=0]{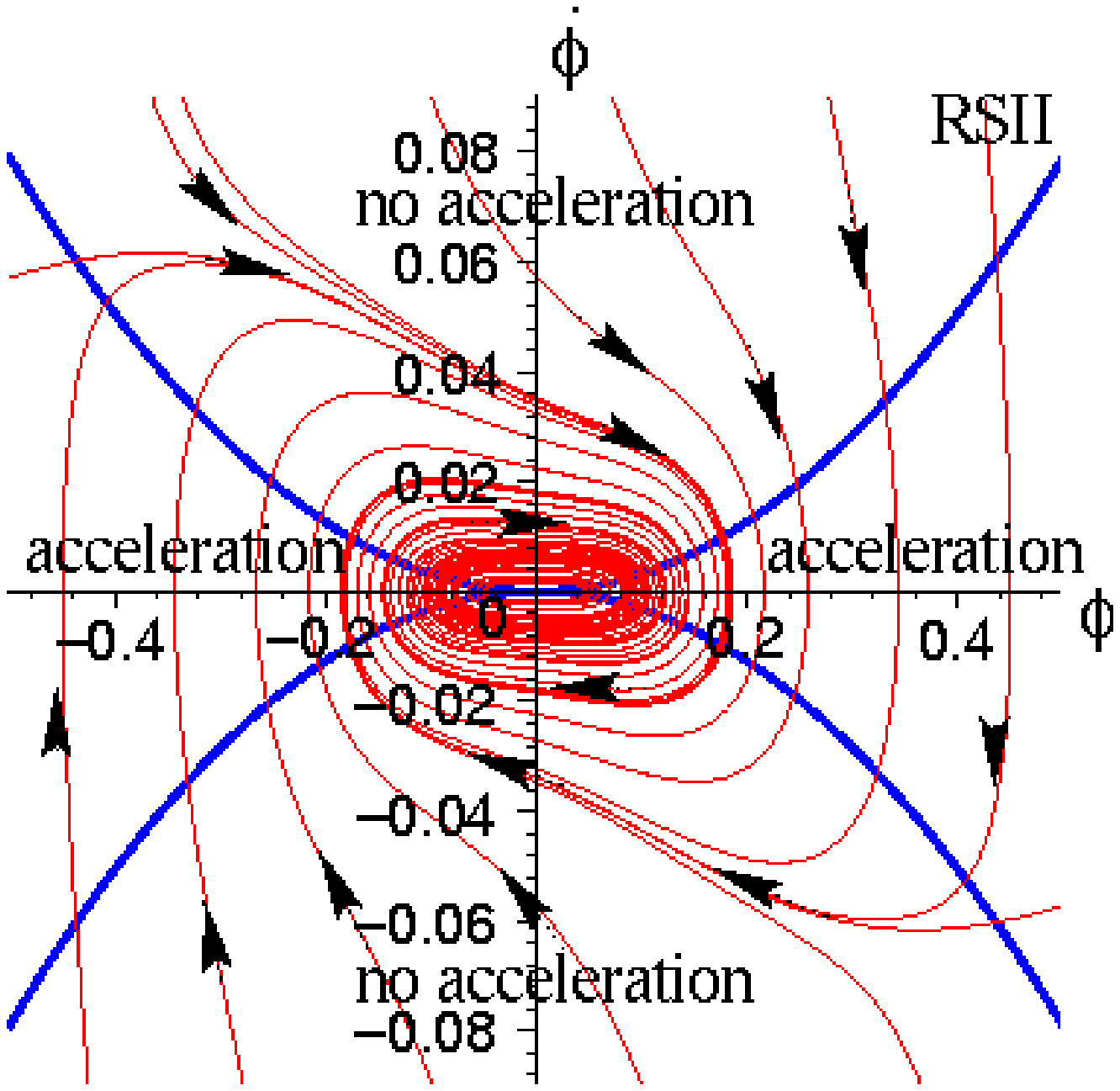}
\\ \vspace{0.7cm}
\includegraphics[width=6cm,height=5cm,angle=0]{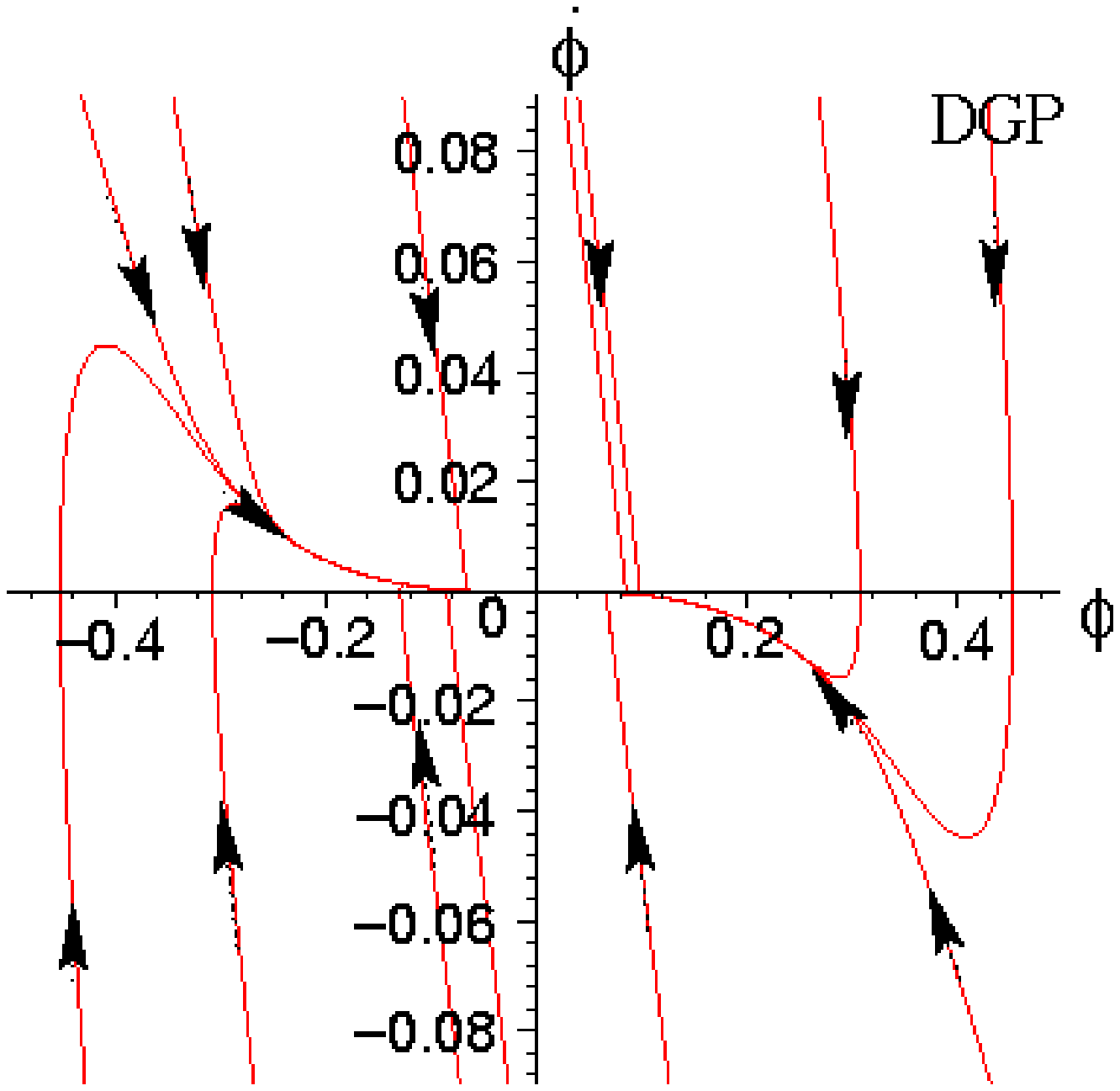}
\end{center} \caption{The phase space of a scalar field with $V(\phi)=\frac{1}{4}\sigma
\phi^4 $ for the standard cosmology, RSII and DGP brane
cosmologies. The solid lines are the acceleration condition.}
\label{f8}\end{figure}

\subsection{Modified exponential potential}
The potential
\begin{equation}
V(\phi) = V_0 \left[\cosh(\frac{\alpha}{M_4} \phi)-1\right]^q
\label{coshpot}
\end{equation}
 is too steep to yield inflation in standard cosmology but in the RSII
braneworld with $q=1$, it can give steep inflation (with large
amplitude gravitational waves). For $q = 1$ the potential has
asymptotic forms $ V(\phi)\simeq V_0 e^{\alpha \phi/M_4}/2 $ when
$\alpha/M_4 \gg 1$, and $V(\phi) \simeq V_0 (\alpha \phi/M_4)^2 /2
$ when $\alpha/M_4 \ll 1$. The exponential potential can not give
oscillating inflaton reheating, but the $\cosh$ potential gives
chaotic inflation behavior \cite{ss}.

 The potential has also been used for quintessence and cold dark matter
models in standard cosmology \cite{sw}. During the oscillation
phase at late times, the mean equation of state is $\langle
w_{\phi} \rangle  =  (q-1)/(q+1)$. The equation of state dictates
that dark energy can be achieved if $q \leq 1/2$ and that to
obtain dust (cold dark matter) we need $q=1$ \cite{turner, sw}.

In Figs. \ref{f4}, \ref{f5} and \ref{f6}, we illustrate phase
diagrams of a scalar field with potential $V(\phi) = V_0
\big[\cosh(\alpha \phi/M_4)-1\big]$, for standard cosmology (where
it plays the role of cold dark matter), RSII (where it could be
regarded as the inflaton field) and DGP($\epsilon=+1$). We vary
the brane tension for the RSII case and vary the crossover scale
for the DGP case. It can be seen that at larger brane tension or
larger crossover scale, the RSII and DGP phase spaces become
similar to the phase space of standard cosmology.

\subsection{Power law potentials}
The power law potentials \cite{linde}
\begin{equation}
 V(\phi) = \frac{1}{2}m^2\phi^2 \;\;\;\;\;\;\;
\mbox{and} \;\;\;\;\;\;\; V(\phi) = \frac{1}{4}\sigma \phi^4
\label{chaotic}
 \end{equation}
  are the simplest
models for inflation that can generate gravitational waves with
large enough magnitude to be observed in the cosmic microwave
background.  The plots in Figs. \ref{f7} and \ref{f8} show the
phase spaces in standard cosmology and the two types of brane
cosmologies. At late times, the RSII trajectories are similar to
the standard case, but they differ markedly from the standard case
at early times. The reverse holds for the DGP trajectories.

\section{Conclusion}
We have investigated the dynamics of a braneworld with induced
gravity, i.e. the DGP model. Unlike the RSII braneworld, which
modifies general relativity at early times and then recovers it at
late times, the DGP braneworld is like general relativity in the
early universe but modifies the dynamics at late times. The
extra-dimensional gravity effect in the DGP ($\epsilon =1$) model
contributes to late-time acceleration, which provides a possible
way to avoid dark energy. We used approximations to solve the DGP
Friedmann equation in both the early and late universe (equations
(\ref{eppos1}), (\ref{epneg1}) and (\ref{early})). The conditions
for acceleration in the DGP model were derived, equations
(\ref{dgpac}), (\ref{s1}) and (\ref{dgpepnegcon}). At early times,
the DGP Friedmann equation approaches the standard general
relativity limit. Modifications appear at late times. When
$\epsilon=+1$, the solutions give accelerating expansion at late
time. On the other hand, when $\epsilon=-1$, the universe expands
slower than standard cosmology. We analyzed the phase planes for a
range of potentials used for inflation, quintessence and cold dark
matter, and compared the dynamics in the DGP case to the RSII and
standard cases. These phase planes show clearly the qualitatively
different behavior of the DGP model. The phase space in the case
of the potential $V(\phi) = V_{0}[\cosh(\alpha \phi/M_4)-1]$, with
various values of brane tension (RSII braneworld) and crossover
scale (DGP braneworld), showed how the standard cosmology scenario
is approached in the limit.

\newpage
{\bf Acknowledgements:} I am grateful to Roy Maartens who
initiated my interest in this topic and for very helpful
discussions. I also thank David Matravers,  Parampreet Singh,
Shinji Tsujikawa and David Wands for useful discussions. I
acknowledge support from the Royal Thai Government Scholarship and
the British-Thai Scholarship Scheme.


\end{document}